\documentclass[numbib]{imaiai}
\usepackage{amssymb,amsmath,mathrsfs, amsfonts, graphicx,theorem,dsfont,yfonts, version, enumitem,stmaryrd}
\usepackage{pst-all}
\usepackage{subcaption} 
\usepackage[english]{babel}
\usepackage{cancel}
\usepackage{bbold}
\usepackage{multirow}
\usepackage{cleveref}
\usepackage{comment}

\bibpunct{[}{]}{,}{n}{,}{;}

\begin{document}
\title{Dense and sparse vertex connectivity in networks}

\author{{%%%% First author details
\sc Mehdi Djellabi}$^*$,\\[2pt]
IRIT, Universit\'e Toulouse 1 Capitole, CNRS,118 Route de Narbonne, 31062 TOULOUSE Cedex 9, France\\
$^*${\email{Corresponding author: mehdi.djellabi@irit.fr}}\\[2pt]
%%%%%%% Second author details
{\sc Bertrand Jouve}\\[2pt]
LISST (UMR5193), CNRS, Universit\'e Toulouse 2, 
5 all\'ees Antonio Machado, 31058 Toulouse Cedex 9, France\\
{bertrand.jouve@cnrs.fr}\\[6pt]
%%%%%%%
{\sc and}\\[6pt]
%%%%%%% Third author details
{\sc Fr\'ed\'eric Amblard} \\[2pt]
IRIT, Universit\'e Toulouse 1 Capitole, CNRS,118 Route de Narbonne, 31062 TOULOUSE Cedex 9, France\\
{frederic.amblard@ut-capitole.fr}}

\maketitle

\begin{abstract}
{The different approaches developed to analyze the structure of complex networks have generated a large number of studies. In the field of social networks at least, studies mainly address the detection and analysis of communities. In this paper, we challenge these approaches and focus on nodes that have meaningful local interactions able to identify the internal organization of communities or the way communities are assembled. We propose an algorithm, ItRich, to identify this type of nodes, based on the decomposition of a graph into successive, less and less dense, layers. Our method is tested on synthetic and real data sets and meshes well with other methods such as community detection or $k$-core decomposition.} 
{Complex networks, Network analysis, Network density, Rich club, Weighted rich club, Communities, k-core decomposition.}
\end{abstract}

\section{Introduction} 
The work initiated in the 2000s, which led to the creation of a network science, has shown that many large real-world networks share similar structural properties. The major two are the ``small world'' phenomenon \cite{wat99} and the power-law distributions of a number of key measures \cite{alb99,fal99,hub99,red98,kum10}. Many other shared patterns have been highlighted and for more in-depth reviewing the readers can refer to \cite{new06,alb02,dor00,new03a}. The first generic network models, produced mainly before the 2000s (for example, Erd\"os-R\'enyi networks or regular grids to name just two extreme examples) are not really adjusted to these large networks because they are largely homogeneous, whereas real world networks are highly heterogeneous. A lot of research has thus been carried out to upgrade them or to design new ones. In that context, precursor works on the configuration model \cite{mor35,bol80}, allowing for random choices of a network with a prescribed distribution of degrees, are back in the forefront \cite{fos18}. Another model proposed by Watts and Strogatz \cite{wat98} is based on a random rewiring of a regular network. Many other models have been published, and some of them are based on the concept of evolving network \cite{bar99, rav03}. 

Nowadays, however, it is widely agreed that these shared 
topological properties are not sufficient to explain the diverse and complex 
architectures of real world networks. In various fields of application, the existence of a modular functional organization is observed or reasonably assumed \cite{you92, spi03, kum99, lus04,kra03}, and assumptions are made about the existence of links between the topological structure of a network and the function of the system.  Concerning complex networks as a whole, a new idea has emerged, that of a natural division of networks into modules, each of which presenting a relative structural-functional homogeneity and whose integration is carried out at the level of the entire network with a relative complexity \cite{sim62,har99,swi90}. To support these views, theoretical results have shown that structural parameters of the modules may account for some network properties \cite{mor19}. In addition, the  dynamic interconnection of the modules can sometimes largely explain the global dynamics of the system and is therefore an important aid in understanding and modelling complex networks \cite{est11}.
Hence, the concept of community has gradually gained momentum due to the widespread acceptance that a community is a subset of nodes dense in connections with respect to the remaining part of the network, even if some other definitions have been suggested \cite{rad04,was94}. Automatic methods for searching  communities in networks have been a very active field of research in the last few years and are often based on the optimization of combinatorial or statistical criteria (\cite{jav18,for16,les10,rad04} to name a few). 
Depending on the calculation methods, communities may overlap or not and the capacity to evaluate the quality of a community organization generated by an algorithm is a real challenge. The current evaluation processes can be classified into two categories: those relying on artificial networks and those relying on well-known real world networks. Artificial networks contain ad hoc built structural modules \cite{hol83,lan08,lan09} and algorithms are tested on their ability to identify these modules. Concerning real world networks, the situation is more complex because the observed modularity is often both functional and structural. Recent works focus on the bias that may result from the confusion between metadata and ground truth for community detection algorithms \cite{hri14}. There is neither universal ground truth nor universal community detection algorithm \cite{sch17} so that  the validation of communities detection algorithms thanks to metadata must ensure that the information conveyed by the metadata is well represented by structural configurations \cite{pee17}.

The recurring observations showing that nodes with similar functions have a higher chance to be linked to one another than random pairs of nodes, have given rise to a number of valuable studies whose basic principle, inspired from statistical methods, is to consider as relevant those configurations that deviate most from a random null model to be defined. 
The modularity measure \cite{new04, new10}, that has received a considerable amount of attention, is based on the above-stated comparison which considers the configuration model as a null model. Other examples can be found in \cite{lan11,bia09}. The choice of the null model is both  very flexible and crucial. Such  flexibility made it clear that very different configurations could be valid and reinforced the existence of a complexity that remains to be explored. The way networks are built impacts the type of null model to be considered, e.g technological and social networks will have to be treated differently  \cite{new03,cre17}. Besides, one  should revisit the fact that a network can simply be partitioned into homogeneous communities and specific  methodologies should be developed in order to bring heterogeneity into and between structural communities, by taking into account, for instance, sparser parts of the network or  the multi-scale aspects of complex networks. 
These points have been already addressed in several recent papers. Some works suggest to differentiate the types of nodes in a  given community according to their position  on the community: either ``centered" or ``relay"  with  other communities \cite{sei13,kel12,wan11}. 
The sets of nodes overlapping several communities have been studied more detainedly, leading to changes in the definition of a community \cite{rad04}. Whereas one could intuitively expect that these sets are less dense in connections, \cite{yan12} shows  the opposite. With more and more data available at different scales, the measures we have on the networks should help to integrate these different levels from local to global. In \cite{jen16}, the authors modify the betweenness centrality measure so as to have two different terms, one depending on the entire network (standard  part of the betweenness) and the other one only depending on the immediate neighborhood of the node. Along the same lines, \cite{che18,xia12} modify the expression of Newman's modularity in order to take onboard the neighborhood configurations and thus solve the limit resolution issue. 
The concept of modularity is also adapted to hierarchical patterns which play a crucial part in the organization of large systems \cite{meu10,sim62,rav02,rav03,lan09a,mon12},

A network without a clear modular organization (for example because many nodes are artificially attached to communities),  may present a ``core/periphery" structure \cite{bor99,cse13,rom14,hol05}. That type of  meso-scale configuration has been less studied than community structures but is complementary \cite{xia18} and somehow more general if one takes into account several core models. A core was originally defined in \cite{bor99} as a set of nodes that are both largely connected to each other and to peripheral nodes, the latter being poorly connected with one another. Several cores may exist, a core may not be in a central position but may occupy a central position for one part of the network. Let us note that a rich club  \cite{col06,zho04} is slightly different from a core and refers to the fact that nodes of higher degree (hubs) are more densely connected than smaller degree nodes. A rich club may exist even if the remaining network does not have the properties inherent to a periphery. 

Our approach is part of the current trend which explores the new facets of topological network structures. Starting from some studies which have shown that between a core and a periphery \cite{ver14}, between a rich club and the rest of the network \cite{bou08}, there may well be  meaningful configurations of nodes and links worth exploring, we have chosen to focus on nodes with relatively few connections in their neighborhoods.
Our considerations are inspired by Burt's work \cite{bur92}, an expert in network sociology who, starting from the  notion of ``structural hole",  establishes bridges between personal networks on the one hand and complete networks on the other hand. So, we define a local weight measurement that takes into account both the degree of a node, and the properties of the edges attached to it. A multi-scale approach is obtained by successive deletions of sets of highly weighted nodes. These sets are kind of rich-clubs \cite{zho04} with respect to this weight measurement. The existence or not of such a rich-club is decided in comparison to a null model to be defined. The nodes subsequent to  all the successive deletions of rich clubs are neither in the cores of communities nor in their overlaps. Such nodes are hardly highlighted by betweenness centrality analyses, although they can be regarded as a kind of backbone supporting the architecture of the network. The method implemented in this study is to be assessed on artificial networks and some well-known real networks. The paper is organized as follows. Section 2 provides a quick overview of the different topological measures that can be attached to an edge and to a node and introduces the $\delta$ measure that will be used. The distributions of $\delta$ on some standard networks will be presented. Then, in Section 3, the rich-club idea of \cite{zho04} will be adapted to a weighted network with topological weights, and the null model relative to $\delta$ will be defined. Section 4 will describe the iterative rich-club extraction algorithm that can be used to assign to each node a measure of the quality of its belonging to the sparse part. Section 5 will summarize the results of the method on synthetic and standard real network models.  Concluding remarks in Section 6 will address some general perspectives.

\section{Topological edge weights, and the $\delta$ measure}
\subsection{Some notations}
The networks are finite, undirected and without any  loop or multiple edges. We designate by $G=(V(G),E(G))$ a network with a node  set $V(G)$ and an edge set $E(G)$.  
By default, $N$ will stand for the number $\vert V(G) \vert$ of nodes of $G$.
 We write $i\sim j$ when $i$ is adjacent to $j$, i.e $(i,j) \in E(G)$. Please note that as $G$ is not looped, we always have $i\not \sim i$.
The (open) neighborhood of  $i$ in $G$ is  $N_G(i)=\{ j \in V(G)\:,\:i \sim j\}$. For any node  $i \in V(G)$, we denote by $d_G(i)=\vert N_G(i) \vert$.  If confusion can be discarded, we note the previous quantities by $d(i)$ for the degree and $N(i)$ for the neighborhood. 
\newline Let us remind that the adjacency matrix $\textsc{A}=(a_{ij})_{1 \leq i,j \leq N}$ of $G$ is defined by $a_{ij}=1$ if $i \sim j$ and $0$ otherwise.
A clique of a network $G$ is a maximal complete subnetwork of $G$, in which all the possible edges do exist. %For $n\geq 1$, the complete network (resp. cycle) with $n$ nodes is denoted by $K_n$ (resp. $C_n$). 
\subsection{Definition of $\delta$}
In line with the latest methods of network analysis, the method explored in this paper is a multiscale method aimed at  representing the network as an arrangement of different local configurations. In order to obtain a finer  description  of the network than its mere division into communities (in the standard  sense of densely connected subsets of nodes), we first consider nodes that are in the ``core" of the communities. This means that most of the edges incident to such a node  remain within the community. We consider that an edge is within a community if both the degrees and the relative number of common neighbors of its extremities are relatively high. Hence, we define the weight $w(i,j)$ of a couple of nodes $i$ and $j$ by:
 $$ w (i,j)= k \cdot d(i) \cdot d(j) \cdot S_{DCS}(N(i),N(j)) \cdot a_{ij}$$
where $k=\frac{1}{(N-1)^2 (N-2)}$ is a normalization factor, $a_{ij}$ is $1$ if $i \sim j$ and $0$ otherwise, 
and
$$S_{DCS}(N(i),N(j))=2\cdot\frac{\vert N(i) \cap N(j) \vert}{d(i) + d(j)}$$
is the
Dice-Czekanowski-S\o rensen \cite{dic45} similarity index. The measure $w$ is then equal to the product of the degrees weighted by a topological overlap measurement.

Local configurations are densely connected clusters whose possible internal heterogeneity will not be touched upon in this paper since several methods are already available.  We will however focus on understanding the structure of the overall arrangement of these configurations. To that end, each node will be weighted by an indicator of its local topology called ``topological strength" and the arrangement will be identified on a weighted network (cf. \Cref{iterative_extraction}).

We define the strength $\delta(i)$ of a node  $i$ by the sum of the weights of its adjacent edges:
\begin{equation}
 %\delta(i)= \frac{1}{N-1} \cdot \sum_{j \in N(i)}w(i,j) 
 \delta(i)=   \sum_{j \in N(i)}w(i,j) 
 \label{(1)} 
 \end{equation}
By convention, $\delta(i)=0$ if $i$ is an isolated node  that is $N(i)=\emptyset$.
It is easy to show that 
\begin{equation}
    w (i,j)= k \cdot\vert N(i) \cap N(j) \vert \cdot H\big(d(i),d(j)\big)\cdot a_{ij}
    \label{(w)} 
\end{equation}
where $H\big(d(i),d(j)\big)$ is the harmonic mean of $d(i)$ and $d(j)$. Let us note that, within the \Cref{(1)}, the harmonic mean makes the contribution of an edge between a node of low degree and a node of high degree low. Moreover, the weight $w (i,j)$ of an edge  is in $[0;\frac{1}{N-1}]$ and is equal to $0$ when nodes $i$ and $j$ have no common neighbors and $w(i,j)=\frac{1}{N-1}$ if, and only if, $i$ and $j$ are neighbors with all other nodes of the graph, ie. $N(i)\setminus \{j\}=N(j)\setminus \{i\} =V \setminus \{i,j\} $ .

%Hence, $\delta(i)=d(i)$ if, and only if, every edge incident to $i$ has a weight equal to $1$ which is equivalent to the case where the graph is a clique . In the general case $0 \leq \delta(i) \leq d(i)$. \vspace{0.2cm} \newline 
Hence, $\delta(i)=1$ if, and only if, every edge incident to $i$ has a weight equal to $\frac{1}{N-1}$ which is equivalent to the case where the graph is a clique . In the general case $0 \leq \delta(i) \leq 1$.  
If we consider the mean edge weight $\overline{w}=\frac{1}{m}\sum_{(i,j) \in E(G)}w(i,j)$, the mean degree  \vspace{0.2cm} \newline $\overline{d}=\frac{1}{N}\sum_{i\in V(G)}d(i)$ and the mean node  weight $\overline{\delta}=\frac{1}{N}\sum_{i\in V(G)}\delta(i)$, we have
$\overline{\delta}=\overline{w}\cdot\overline{d}$.

\subsection{Distribution of the $\delta$ values for some selected networks}\label{distribution of delta}

Graph metrology has produced many measurement indices that apply to nodes and some indices applying to edges \cite{new10a,qua17}. The degree centrality and the clustering coefficient measures are  particularly essential  in network analysis and closely related to the definition of $\delta$. With a view to illustrate the difference between these two measures and $\delta$, we compare their distributions for two different  types of networks: a network produced by a Watts and Strogatz random model \cite{wat98} and a network produced by a stochastic bloc-model (i.e. SBM model) \cite{hol83}. 
\begin{figure}
    \centering
    \begin{subfigure}[b]{0.45\textwidth} % "0.45" donne ici la largeur de l'image
        \centering \includegraphics[width=\textwidth]{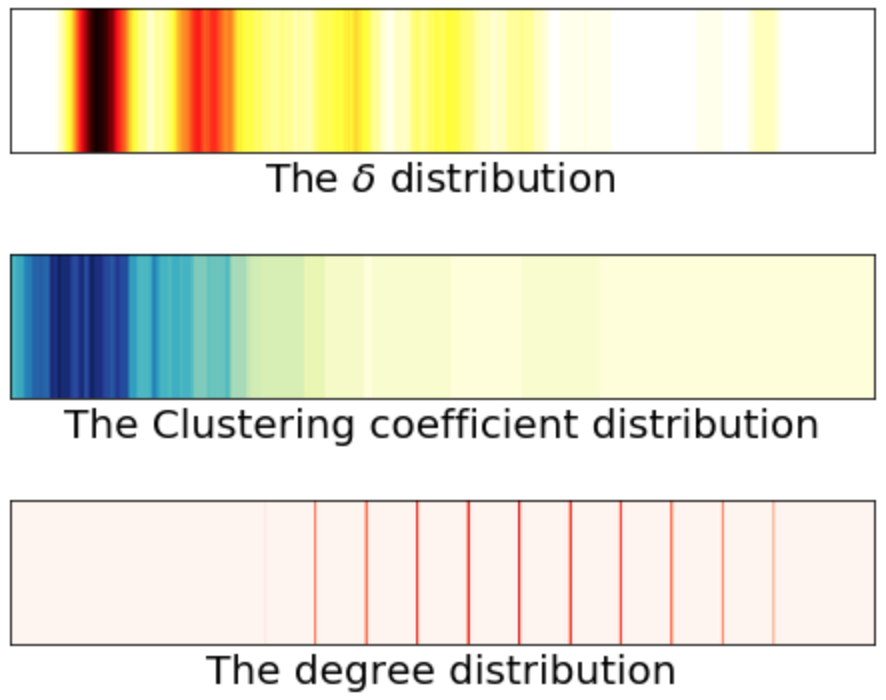}
        \caption{Heat scale for the Watts and Strogatz model.}\label{distribution-a}
    \end{subfigure}
    ~ % ce symbole ajoute un espacement horisontal entre les premiÃ¨res deux images
    \begin{subfigure}[b]{0.45\textwidth}
        \centering \includegraphics[width=\textwidth]{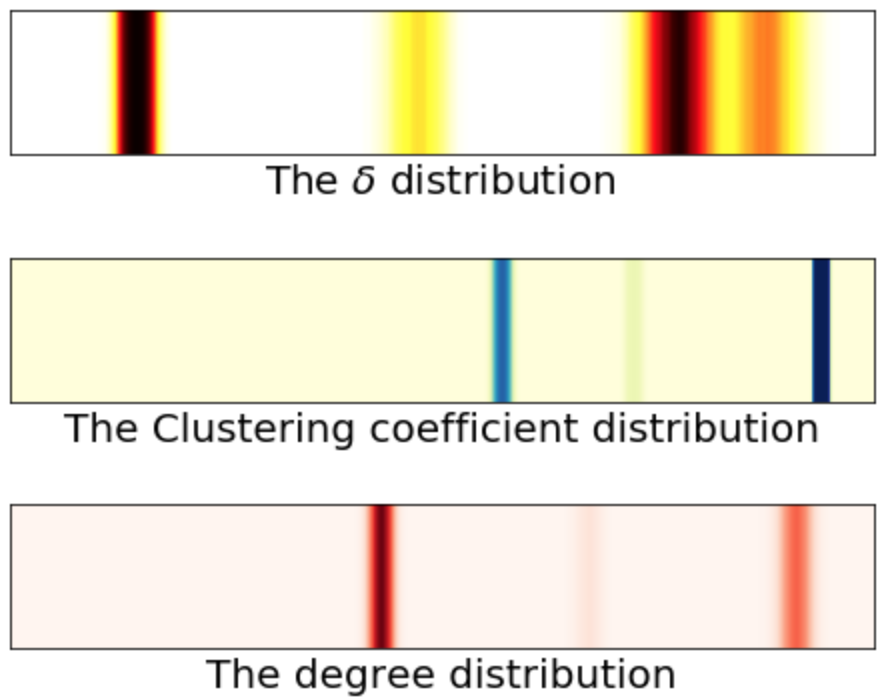}
        \caption{Heat scale for the block stochastic model.}\label{distribution-b}
    \end{subfigure}

    % la ligne blanche correspond au retour Ã  la ligne aprÃ¨s le deuxiÃ¨me image
    \begin{subfigure}[b]{0.44\textwidth}
        \centering \includegraphics[width=\textwidth]{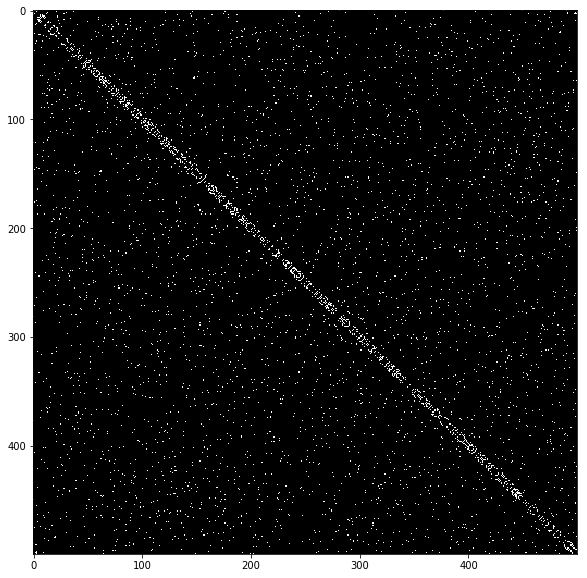}
        \caption{Adjacency matrix for the Watts and Strogatz model. }\label{distribution-c}
    \end{subfigure}
    ~ 
    \begin{subfigure}[b]{0.45\textwidth}
        \centering \includegraphics[width=\textwidth]{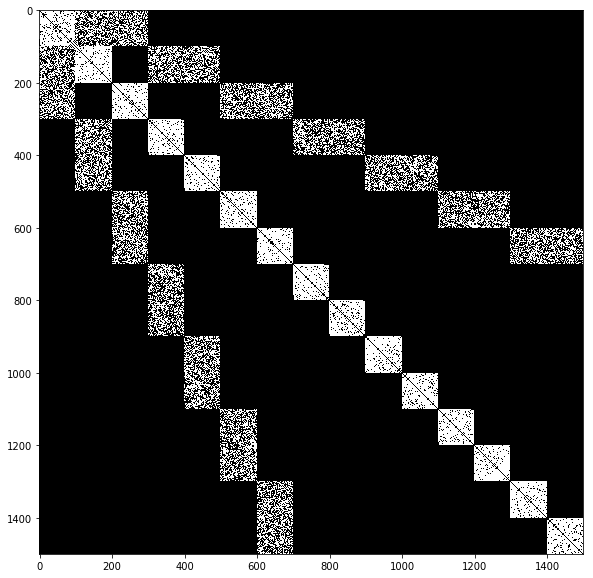}
        \caption{Adjacency matrix for the block stochastic model.  }\label{distribution-d}
    \end{subfigure}
    \caption{Adjacency matrices and heat scales representing the distributions of $\delta$, degree and clustering coefficients for: Watts-Strogatz model (a,c) and Stochastic Block-Model (b,d).}
    \label{distribution}
\end{figure}

The SBM model (or planted partition model) used in this study consists in 15 blocks connected by a tree-like architecture. Each block contains 100 nodes with an internal probability $p=0.85$, whereas the nodes of the ``neighboring" blocks are connected with a probability $q=0.65$ (\Cref{distribution-d}). The model is constructed so that the architecture connecting the blocks together is  of rooted tree type, all its leaves being at a distance of 3 from the root. The root is of degree 2 and all other internal nodes have a degree 3, with parameters $p=0.85$ and $q=0.65$, as can be seen in \Cref{distribution-d}. Each block is characterized by given average values of $\delta$, degree and clustering coefficient. It is thus possible to use these average values to differentiate the  various types of blocks. 
For example, in \Cref{distribution-b}, each  distribution of the degree and of the clustering coefficient has 3 modes, which corresponds to 3 types of blocks:  the root block, the internal blocks and the leaves blocks. Yet,  4 modes for the $\delta$ distribution can be observed: two corresponding to the root block and to the leaves blocks and two corresponding to the internal blocks, one type of block being  linked to the root (two blocks) and one type being linked to the leaves (four blocks). 
This distinction shows that the nodes inside the internal blocks have on average the same degree and clustering coefficient, but a different $\delta$, depending on their neighboring blocks. This is due to the fact that the $\delta$ measure takes into account the degree of each block (hereby, we mean by degree of a block $C_i$ the number of other blocks $C_j$ in which the probability that a node in the block $C_i$ is linked to a node in the block $C_j$ is not equal to $0$) as well as the degree of its neighboring blocks, while the average clustering coefficient within a given block is just expressed as a function of its degree. The formal expressions of the $\delta$ measure, and the clustering coefficient are stated in Appendix.  

The Watts-Strogatz model is made of 500 nodes  designed  from a regular ring lattice on which  each node  is linked to the 10 nearest neighbors and with a rewiring probability $p=0.75$ of the edges (\Cref{distribution-c}). Contrary  to the  stochastic block model, the Watts-Strogatz model makes it  more difficult to identify the different types of nodes although  the $\delta$ distribution  evidences two prevailing modes as well as several other modes of lesser magnitude, that cannot be traced when observing the  distribution of the clustering coefficient. As for the degree distribution, it does not contain enough information, because all the nodes have a degree that is distributed around the initial value, namely $d=10$, with a slight variance due to randomization. 

For both random models,  \Cref{distribution} shows that the distribution of $\delta$ is different from the other  two distributions and may  contain more or different information.
\section{ Weight $\delta$  in the context of the rich club phenomenon}
\subsection{Rich club phenomenon and existing null models}
The rich club phenomenon refers to a topological effect in complex networks, which can be defined as  the tendency of high degree nodes (rich nodes) to be interconnected, thus forming a a  rich club subset which  was first formalized in a quantitative form in \cite{zho04}.  An algorithm for automatic detection  was then  proposed in \cite{col06}.

The rich club effect is quantified by the following parameter $\phi(k)$:
\begin{equation}
    \phi(k)=\frac{2\cdot E_{>k}}{N_{>k}\cdot(N_{>k}-1)}
    \label{phik}
\end{equation} 
where $E_{>k}$ is the number of edges among the $N_{>k}$ nodes having degree higher than $k$. Thus, $\phi(k)$ is the edge density of the subnetwork induced by nodes with degrees greater than $k$. For a network with the above-mentioned rich club properties the $\phi$ curve may have a peak around a given value of $k$. In order to avoid the bias due to the fact that high degrees are more likely to be linked together, \cite{col06} proposed a normalized version of $\phi$: 
\begin{equation}
    \rho(k)=\frac{\phi(k)}{\phi_{null}(k)}
    \label{rhok}
\end{equation} 
where $\phi_{null}(k)$ is the rich club coefficient of a null model. 
The rich club is then obtained for the value of $k$ such that $\phi(k)$ is significantly higher than $\phi_{null}(k)$, i.e. when $\rho(k)$ is maximum. 
For unweighted networks, the null model preserves the degree distribution and breaks the degree-degree correlations between neighboring nodes. The algorithm presented in \cite{col06} is widely used in the case of undirected unweighted networks, and consists in choosing the configuration model as a null model, by preserving the degree sequence of the original network, while randomly rewiring the endpoints of each edge.
In the case of weighted networks, however, there is no consensus on the rich club coefficient nor on the choice of a null model, even if \cite{als14} proposes a unified framework. Indeed, the degree distribution is no longer the only quantity that should be kept in the null model. Ideally one should  preserve the distribution of weights, and the distribution of strengths, which is the sum of the weights of all the edges attached to a node. Finding an algorithm that provides a null model meeting  all these requirements is not an easy task.

In \cite{zla09}, the authors construct a null model with a given strength distribution, but change both the weight and the degree sequences.
In \cite{ser08}, the rich club coefficient is calculated by replacing the degree of a node with the sum of the weights of its incident edges and by normalizing the result using one of the functions listed in \cite{als14}. The authors of \cite{ops08} propose  a randomized directed network as a null model by replacing each non-directed edge with two directed edges, one in each direction, and by reshuffling the endpoints of the outgoing edges of each node among its neighbors.
Although this null model preserves the degree, weight and strength sequences, its main limitation is that it does not break the degree-degree correlations. This becomes a major issue when the strength is correlated with the degree because, in such a case, the null model will also have a strength-strength correlation. 
Another null model for weighted networks is proposed in \cite{serrano06}. It generates a network with degree and strength distributions that converge towards some given distributions when the network is large enough.
We have made empirical observations on real world networks suggesting that this size limitation has a greater impact  when the studied networks are characterized by heavy tailed degree and/or strength distributions.
\subsection{An adapted ``null model" for weighted networks}
Given a network $G$, we consider the weighted network $W_G$ with adjacency matrix $\textsc{W}=(w(i,j))_{1 \leq i,j \leq N}$. 
So, $W_G$ is an auxiliary network with weights equal to $w(i,j)$, in such a way that the sum of the weights of the edges attached to a node $i$ is equal to $\delta(i)$.

It is a priority for our study to choose a null model that does not contain any $\delta$-$\delta$ or degree-degree correlations between neighboring nodes, in order to ensure that the rich nodes (i.e. high $\delta$ nodes) are linked together only by chance. But this model must also preserve the degree and the weight distributions. 
The null model applied in this study has been extensively studied in \cite{LiuXu}, and is provided by the following algorithm: 
\begin{enumerate}
    \item Starting from $G$, we first modify the topology by randomly rewiring its links, in order to break the degree-degree correlations, and thus obtain a configurational model of $G$ that we call $G_{null}$, which no longer contains degree-degree correlations. 
    \item Then, the weights of the network $W_G$ are randomly redistributed on the edges of $G_{null}$ which yields a null model, noted $W_{G,null}$, that has neither degree-degree nor $\delta$-$\delta$ correlations. 
\end{enumerate}

By construction, this null model preserves the degree and the weight distributions, and therefore the average value of $\delta$, but does not preserve the distribution of $\delta$. This involves a modification in the computation of the rich club coefficient, that is detailed in the next section. 
\section{An iterative extraction algorithm}\label{iterative_extraction}
\subsection{A new approach for weighted rich clubs}
In this section we present an algorithm for the extraction of the high weighted subsets of nodes, using a weighted rich club approach. 

Since the null model does not preserve the $\delta$ sequence computed on the original network, we need to filter out the nodes following a new strategy, different from the one that consists in selecting the nodes whose strength is larger than a certain value $\delta$. 

So, given a network $G$, let us denote by $V_n$ the set consisting of  $n$ nodes \footnote{If several nodes have the same $\delta$, we randomly choose the order in which they are added. In practice, this situation is very unlikely and the choice order has a negligible impact in large real networks. } of highest $\delta$ 
\begin{equation*}
    V_{n} =\{u_1,u_2,...,u_n \mid \delta(u_i)>\delta(u_{i+1})\}
\end{equation*}
and define the weighted rich club coefficient $\phi(n)$ in the following way:
\begin{equation}
    \phi_G(n) = \frac{\sum\limits_{(i,j) \in E\cap (V_n\times V_n)} w(i,j)}{\sum\limits_{(i,j) \in E} w(i,j)}.
\end{equation}

Comparing this weighted rich club coefficient to that obtained from the null model (by analogy with the case of unweighted networks) this definition of $\phi_G(n)$ ensures that the two quantities are calculated from networks with the same number of nodes $n$, as it is the case in  \Cref{phik}.

Now we define our weighted rich club parameter $\rho_G(n)$ as 

\begin{equation}
    \rho_G(n) = \phi_G(n) - \phi_G^{null}(n)
\end{equation}
where $\phi_G^{null}(n)$ is the weighted rich club coefficient computed from the null model. We then define a high weighted subset of nodes in $G$, with respect to the definition $\delta$ of density, as the set of nodes that maximizes the weighted rich club parameter:
\begin{equation}
V_M \; : \; M = \text{arg max}_n\{\rho_G(n)\}
\label{VM}
\end{equation}
Compared to the classical method for unweighted networks, note that we consider the difference $\phi_G-\phi_G^{null}$ instead of the ratio $\phi_G/ \phi_G^{null}$. This choice is justified when we have ${\phi_G(n)}\ll 1$ and  ${\phi_G^{null}(n)}\ll 1$, and at the same time a high value for the ratio $\phi_G(n)/ \phi_G^{null}(n)$. This case occurs for small values of $n$ and is encountered for networks whose auxiliary model $W_G$ contains a strong $\delta$-$\delta$ correlation among neighboring nodes, while having a heavy tailed $\delta$ distribution. This is due to the presence of a small number of nodes that have values of $\delta$ much higher than the average, and that are also linked to one another, as emphasized by the $\delta$-$\delta$ correlations. The consequence of such a configuration is both a rapid growth of ${\phi_G(n)}$ when $n$ is small and a slow growth of $\phi_G^{null}(n)$ due to the randomization process. As a consequence, the ratio $\phi_G(n)/ \phi_G^{null}(n)$ reaches its maximum for small values of $n$, which results in a small weighted rich club, sometimes containing only two nodes and one edge, which is not very relevant to us\footnote{ Another justification for this choice is practical in regard to the iterative algorithm we use, which takes significantly less calculation time when we evaluate the difference than the ratio}.

\subsection{The shortcomings of a single execution step}\label{toy_model}
 Let us  highlight the fact that a single execution of the above algorithm is not sufficient  to provide a full information about the overall structure of a network. So,  we introduce a toy model of network made of several blocks of the same size, each block being an Erd\"os-R\'enyi network with different parameters. Then we connect the nodes of the tree to the blocks with a certain probability. 
  
  The parameters of this model are:
\begin{itemize}
\item $N_b$: the number of blocks $C_i$ in the network  
\item $N_i$: the number of nodes inside each block $C_i$, $i\in \{1,2,..,N_b\}$,
\item $P_i$: the probability to create a link between any two nodes inside the same block $C_i$,
\item $N_t$: the number of nodes contained in the random tree,
\item $p^t_i$ the probability to connect any node of the tree to any node of the block $C_i$. 
\end{itemize}
This generates a network with $\sum_{i=1}^{N_b} N_i +N_t$ nodes. We first generate a random tree as the spanning tree of a Erd\"os-R\'enyi random network whose number of nodes is $N_t$. Then, we generate the $N_b$ blocks independently and connect each node of the tree to each node of the block $C_i$ with the probability $p^t_i$.

To ensure that the toy network generated in this way is connected, it is sufficient to choose $p^t_i$ so that the average number $M$ of links connecting the tree to the blocks is very large compared to 1. It is easy to verify that $M= \sum_{i=1}^{N_b} N_t\cdot N_i \cdot p^t_i$ and then $M \gg 1$ if $p^t_i \gg \frac{1}{N_t \cdot N_i}$. 

In the following, we give an example and a graphical representation (using a spring force positioning algorithm) of the described toy network, with $N_c=4$ blocks of $N_b=50$ nodes in each block, a tree of $N_t=100$ nodes, and internal probabilities: $P_1=0.8,\, P_2=0.6,\, P_3=0.4,\, P_4=0.2$. Here we take $p^t_i = \frac{N_i}{N_t \cdot \sum_j N_j}$, which fulfills the connectivity prerequisite.

\begin{figure}
  \begin{subfigure}[b]{0.62\textwidth}
    \includegraphics[width=\textwidth]{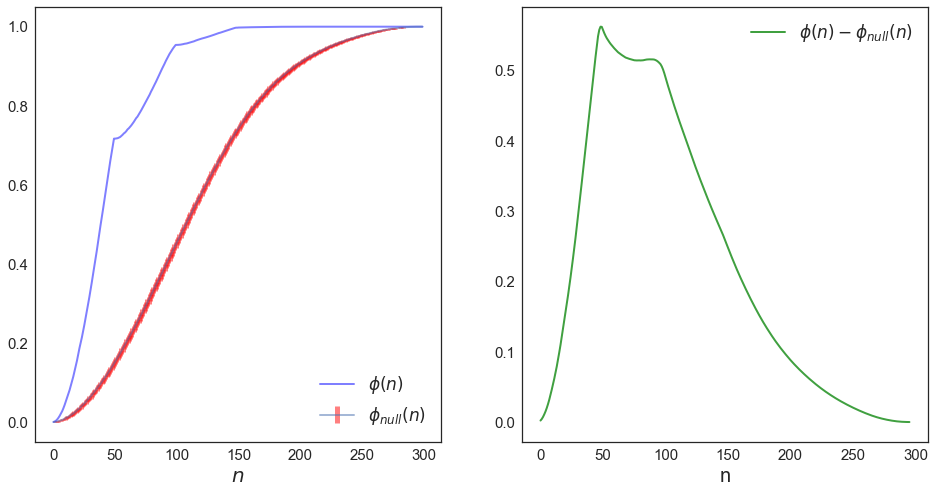}
    \caption{}
    \label{fig:1}
  \end{subfigure}
   % \begin{center}
  \begin{subfigure}[b]{0.35\textwidth}
    \includegraphics[width=\textwidth]{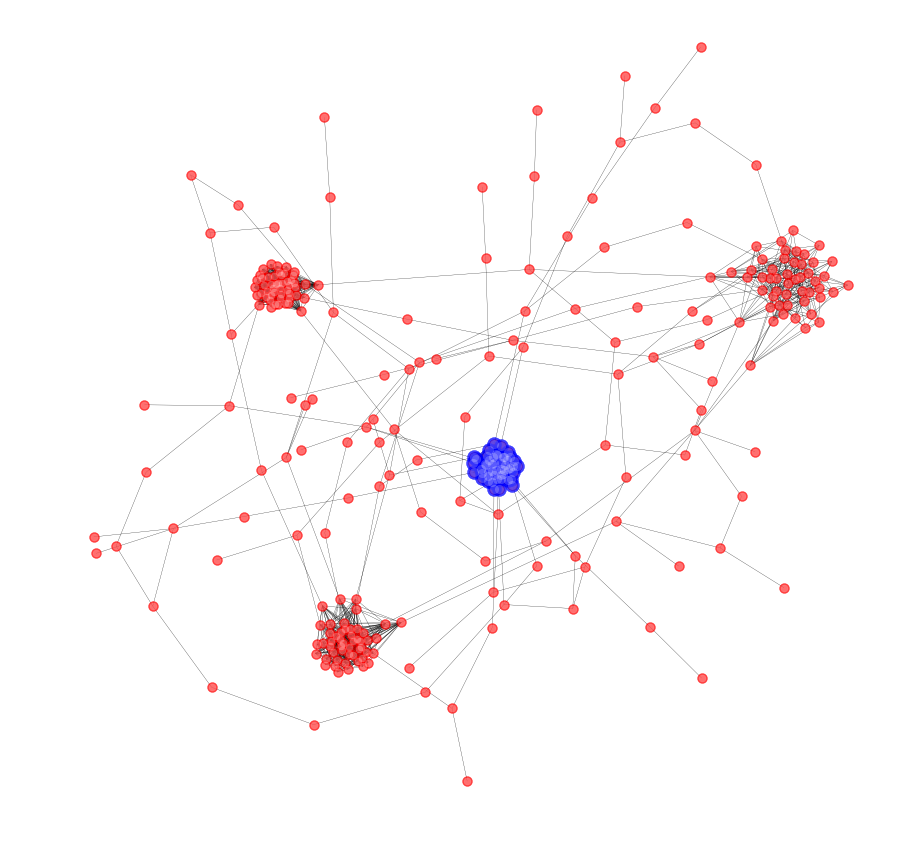}
    \caption{}
    \label{fig:2}
  \end{subfigure} 
%\end{center}
\caption{(a) The evolution of the weighted rich club parameters and the weighted rich club coefficient for both the toy network and its corresponding null model (see \Cref{toy_model} for definitions of the toy network model). (b) The result of a single iteration algorithm. The nodes detected as the dense subset are in blue and represent only one of four dense parts. }
\label{coef}
\end{figure}

Once the toy network has been generated, the procedure described in the previous section is applied; the  results are presented in \Cref{coef}. The maximum of $\rho(n)$ is reached for $n=50$ and the corresponding nodes are those of the cluster with the probability $P_1=0.8$. 
This example shows that it is necessary to iterate the algorithm more than once otherwise nodes that are in other clusters may be counted in the sparse part.
\subsection{Iterative algorithm and quality measure}
We repeat the above-described  process iteratively, while deleting at each iteration $i$, the weighted rich club calculated using \Cref{VM} at iteration $i-1$, and keeping the same weights on the remaining links after deletion. This makes it possible to extract the high weighted subsets of nodes one by one, and this process is repeated until the stopping criterion of the algorithm is reached or there is no more weight in the remaining network. Once these iterations are completed, we obtain a series of weighted rich clubs and the $i^{th}$ element can be evaluated using the difference between the weighted rich club coefficient on the studied network and its null model, normalized so that it is always lower than 1:
\begin{equation}
    Q_i= \frac{1}{N_i}\cdot \sum\limits_{n=1}^{N_i} \rho_{G_i}(n)
    \label{Q_i}
    \end{equation}
where $G_i$ is the weighted network obtained after deleting  the $i-1$ first weighted rich clubs from $G$ and $N_i$ is the number of nodes of $G_i$. Let us remind that the weights of $G_i$ are not recalculated but inherited from $G$. 

This measure represents the average value of $\rho_{G_i}$ over all possible series of nodes of $G_i$ selected in decreasing order of $\delta$.  The measure $Q_i$ generally goes as follows: decreasing from its maximum value, which is the quality of the first extracted weighted rich club, to lower values referring to the qualities of weighted rich clubs with lower ranks. This decrease is valid for networks having high weighted subsets of nodes, the  measure $Q_i$  however, may otherwise exhibit unpredictable behavior \footnote{For example, if the network does not have a particular structure as in the case of an Erd\"os-R\'enyi network}. In the example of the toy model described in \Cref{toy_model}, the quality measure falls to (almost) zero once all its clusters have been extracted, because not only will the weight of the links that remain in the network (after cluster extraction) be low, but there will also be no more clusters left in the network. This entails the decrease of $\rho(n)$ values and of the $Q$  quality measure. 

The measure $Q_i$ is used in order to accept or reject the $i^{th}$ weighted rich club to be part of the dense part: it is accepted if, and only if, $Q_i>Q_{threshold}$. So, $Q_{threshold}$  is a resolution parameter of the dense/sparse parts that has to be chosen. The higher the threshold, the smaller the dense part and the larger the sparse part. It can be constant during the whole process or recalculated at each step of the algorithm. The best choice of $Q_{threshold}$ varies according to the type of data being studied but, in practice, a value proportional to $Q_1$ is often a good choice. In the following, for the real networks we study, we set $Q_{threshold}=Q_1/10$ as a default value. If we have any a priori information on the structure of the dense (or sparse) part, we can choose a more appropriate threshold to make the method more efficient. For example, for the artificial networks we consider below, the dense part has a modular organization that we can more efficiently retrieve using a variable threshold based on an Erd\"os-R\'enyi model.  

\begin{figure}[!h]
    \centering
    \includegraphics[width=15cm]{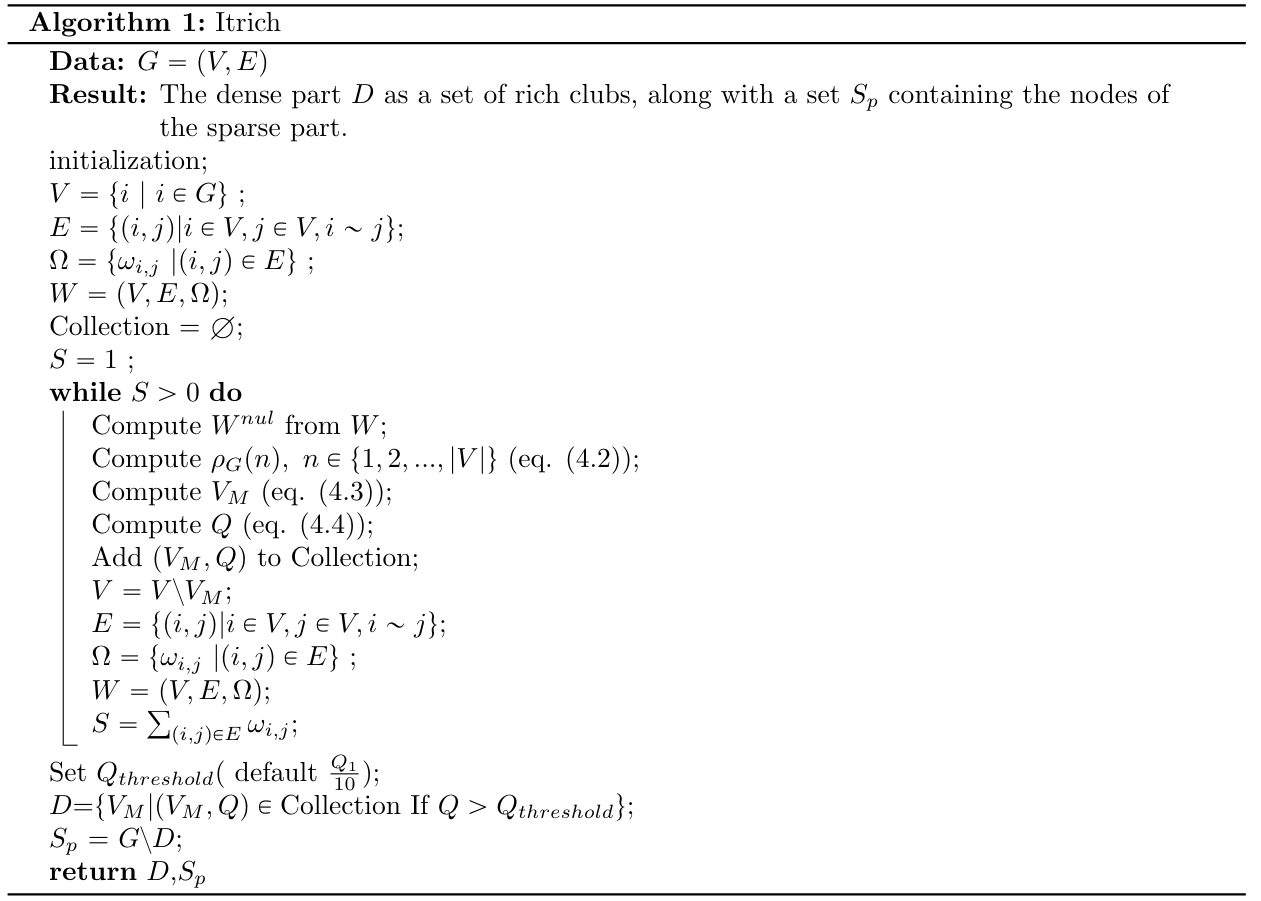}
    \caption{ItRich algorithm}\label{algo1}
\end{figure}

The whole process is summarized in an algorithm named ItRich for iterative weighted rich clubs (\Cref{algo1}).

The complexity of ItRich is proportional to the number of times the main loop is executed. It is not necessary to repeat the computation of the loop until the remaining network has a null weight, if the value of $Q_{threshold}$ is set beforehand (ex: $\frac{Q_1}{10}$). Note that the higher the threshold, the shorter the computation time. The main loop contains 4 computations whose first two can easily be made in parallel. These are the most expensive. The first one consists in calculating a null model. This is bounded by the calculation of the configuration model, which can be obtained in $O(m \cdot \log(m))$ with $m$ the number of links in the network. In practice, it is better to calculate a number of null models and estimate the results based on their averages (as it is made in \Cref{LFR} and \Cref{real-network}).  The second one consists in sorting the list of nodes in decreasing order of $\delta$ (it takes $O(N \cdot \log(N))$), then, in parallel, constructing the $N$ sub-networks induced by $V_n \times V_n$  for $n\in \{1,\cdots,N\}$ and evaluating their weights ($O(m)$ for each sub-network). 
If the decomposition stops after the finding of $N_{rc}$ rich clubs and $m>>N$, then ItRich has order of  $O(N_{rc} \cdot m \cdot \log(m))$ time complexity. Let us note that the computing of the null model is the expensive part of the algorithm.
\section{Results and discussion}
\subsection{Artificial networks}\label{LFR}

We first apply our algorithm to a set of synthetic networks partly used in \cite{lan11}.  
We measure the output of our algorithm ItRich by calculating the values of  Recall and Specificity (also called True Positive rate and True Negative rate respectively). Finally, we compare these performances with those obtained by OSLOM \cite{lan11}, an algorithm widely used in this field. 

For these data, we choose a variable threshold $Q_{ER_i}$ equal to the quality measure obtained on a Erd\"os-R\'enyi random network with the same number of vertices as in the remaining network at step $i$ of the algorithm ItRich (i.e. the initial graph where the $i-1$ first rich clubs have been removed) and a probability parameter $p_i=\overline{\delta}_i^{\frac{1}{4}}$. This value of $p_i$ ensures that the Erd\"os-R\'enyi network and the network evaluated at the iteration $i$ have the same value of $\overline{\delta}_i$ (see Appendix).
As previously noted, the advantage of using a Erd\'os-R\'enyi network is, on the one hand, its lack of modular structure, which ensures that it does not contain any subsets whose density of links is significantly higher compared to the rest of the network. On the other hand it is related to the synthetic data studied in this paper. As seen below, the construction of the data is such that the subnetwork induced by the sparse part is close to a Erd\'os-R\'enyi network.

\vspace{.3cm}
\noindent {\bf Experimental data}

\noindent Our synthetic model is based on a Lancichinetti-Fortunato-Radicchi (LFR) benchmark \cite{lan08} to which nodes have been added as detailed below. Let us remind that in a LFR benchmark network each node is assigned to a community with a control on some characteristics, such as the exponent $\gamma$ of the power-law distribution of the degrees, the exponent $\beta$ for the power-law distribution of the community size, and $\mu$ the proportion of links a node shares with nodes outside its own community. 

Given a LFR, we add nodes so as to create two classes: a first one, called the ``dense part", composed of nodes from the original LFR model, and a second one, the ``sparse part", containing the nodes that have been added. These added nodes will be connected so that their weights $\delta$ are lower than those of the LFR network. This \textit{a priori} classification is then used as a ground truth to evaluate the efficiency of our algorithm in separating the ``dense part" from the ``sparse part".

\begin{figure}
    \centering
    \begin{subfigure}[b]{0.35\textwidth} % "0.45" donne ici la largeur de l'image
        \centering \includegraphics[width=\textwidth]{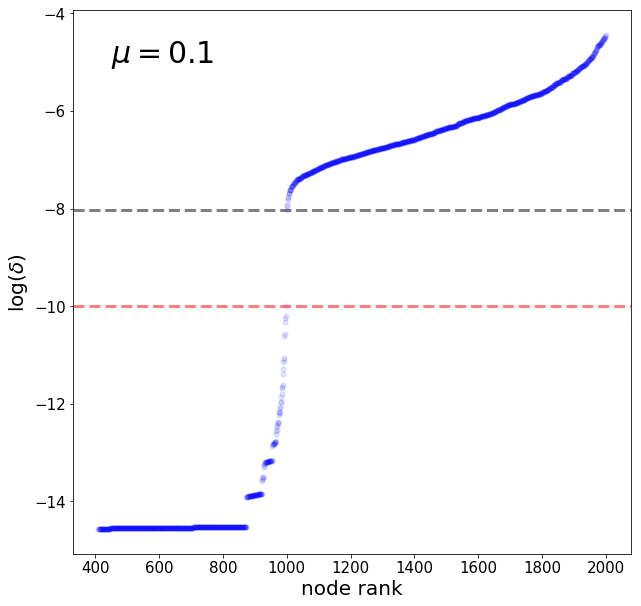}
        \caption{$r=0$}\label{challenge-a}
    \end{subfigure}
    ~ % ce symbole ajoute un espacement horisontal entre les premiÃ¨res deux images
    \begin{subfigure}[b]{0.35\textwidth}
        \centering \includegraphics[width=\textwidth]{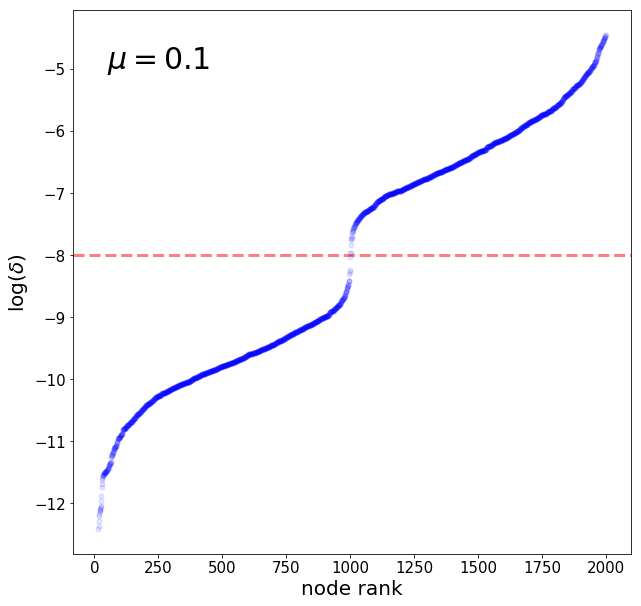}
        \caption{$r=1$}\label{challenge-b}
    \end{subfigure}

\caption{The logarithm of $\delta$ of each node \textit{vs.} its rank. The red dashed lines represent the maximum value of $\delta$ in the noise nodes, whereas the black dashed lines represent the minimum value of $\delta$ in the original LFR ($\gamma=3, \beta=2$).
(a) When $r=0$, there is a gap in the values of $\delta$ separating the nodes of the initial LFR from the noise nodes. (b) When $r=1$, this gap is bridged by adding links between added nodes, so that the minimum value of $\delta$ in the initial LFR network is also equal to the maximum value of $\delta$ in the sparse part.}
\label{challenge}
\end{figure}

In order to connect the added nodes (the noise) to the original LFR network, we use the method of \cite{lan11}: the degree of a new added vertex is drawn from a distribution that is the same as that of the LFR network, and the vertex is connected to the network by preferential attachment. \Cref{challenge-a} shows the ordered values of the logarithm of $\delta$, obtained from an LFR model of 1000 nodes ($\gamma = 3$, $\beta=2$, $\mu =0.1$) to which 1000 other nodes have been added. 

We can observe that the average value of $\delta$ restricted to the nodes of the sparse part (under the red dashed line) is much lower than that restricted to the dense part (above the black dashed line), making the classification quite simple. Indeed \Cref{challenge-a} shows that there is a gap $g_0=\min(\delta_{LFR})-\max(\delta_{noise})$ between the minimum value of $\delta$ in the LFR network and the maximum value of $\delta$ in the set of the added nodes. 
To make the classification more challenging, we randomly add links between the added nodes, which has the effect of reducing the value of the gap. The new gap $g(r) = g_0 \cdot (1-r)$ is given as a function of the initial gap $g_0$ and a parameter $r \in [0;1]$.  When $r=0$ no links are added, and when $r=1$ the initial gap $g_0$ is entirely filled. \Cref{challenge-b} gives the example of a total gap reduction.

In \Cref{challenge}, we can observe that the part of the curve above the red dashed line is the same before and after filling the gap, whereas this is not the case for the noise nodes (lower part of the curve), meaning that the added links change the shape of the $\delta$ distribution of the nodes in the sparse part, rather than only shifting it. This effect is due to the fact that links are added between randomly selected pairs of nodes, and not by especially targeting those with high $\delta$ values, which produces a data set for which the separation between the dense and the sparse part is less obvious to determine. One can point out that the larger the value of $g_0$, the more severely the distribution is impacted when $r$ increases.

 %One can however observe lower levels of perturbations, depending on the parameters used to generate the initial LFR model, which may result in a distribution with a low value of $g_0$ that may be reduced by adding a relatively small number of links to the network. This has a direct impact on the performances of our algorithm as described in the next section. 

%\begin{figure}[!h]
 % \begin{subfigure}[b]{0.5\textwidth}
 %   \includegraphics[width=\textwidth]{rank_stat.png}
 %   \caption{}
 %   \label{fig:1}
 % \end{subfigure}
  %
 % \begin{subfigure}[b]{0.5\textwidth}
 %   \includegraphics[width=\textwidth]{stat_rank_1.png}
 %   \caption{}
 %   \label{fig:2}
 % \end{subfigure}

\vspace{.2cm}
\noindent {\bf Experimental protocol} \label{def_diverses}

\noindent Three experiments are performed, each with a different value of the mixing parameter $\mu$. For each experiment, we perform $N+1=11$ sets of calculations, each set for a different value of  $r$, from $r=0$ to $r=1$. 
For each of the $N+1$ values of $r$, the initial LFR networks have $N_{LFR}=1000$ nodes, $\gamma = 3 $ and $\beta = 2 $.
We execute our algorithm for $N_{noise}$ ranging from $N_{noise}=0$ to $N_{noise}=1000$, in 10 knot increments. 
The algorithm is thus applied 100 times for each $N+1$ value of $r$. Each calculation is moreover based on an average of 50 different null models. 

Let $D$ be the set of nodes from the initial LFR model, $ND$ the set of added nodes, $D_r$ (resp. $ND_r$) the set of nodes detected as the dense (resp. sparse) part by the algorithm, we use the standard following metrics to evaluate the performance of the results:
\begin{itemize}
    \item Recall (or True positive rate): $Sn= \frac{|D \cap D_r|}{|D|}$
    \item Specificity (or True negative rate): $Sp=\frac{|ND \cap ND_r|}{|ND|}$
 %   \item F-score: $F1 = \frac{|D \cap D_r|}{|D|+ |D_r|}$
\end{itemize}

The average values of $Sp$ and $Sn$ over the $100$ networks generated for the different values of $r$ are calculated for the three experiments. The results are then compared with those of the OSLOM algorithm \cite{lan11}. OSLOM is one of the few algorithms that rely on the statistical properties of clusters, while bringing an « added value »  to standard community detection algorithms, i.e the possibility to have a set of ``homeless" nodes assigned to no community whatsoever.  We will consider these nodes as noise, and compare them to the nodes of the sparse part obtained by ItRich. 

\vspace{.3cm}

\noindent \textbf {Results}\\
Only the results for $\mu=0.1$ and $r=0$ and $1$ are plotted (\Cref{ItRichvsOSLOM}) and discussed here and more complete graphics can be found in Appendix 2.
\begin{figure}
  \begin{subfigure}[b]{0.23\textwidth}
    \includegraphics[width=\textwidth]{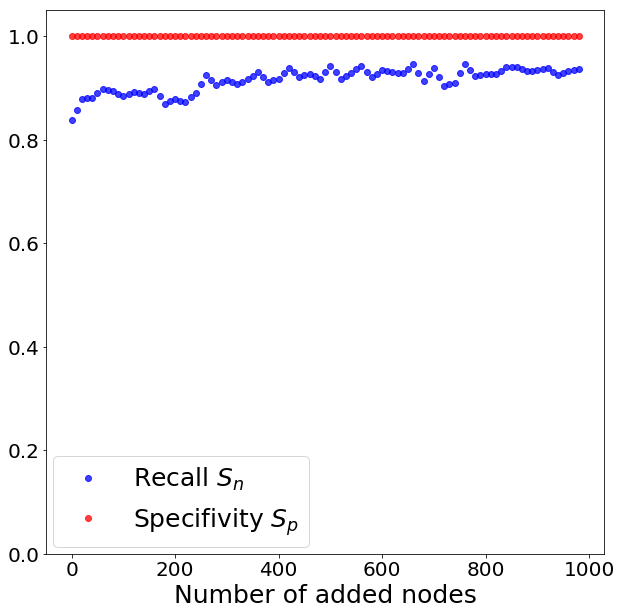}
    \caption{ItRich $r=0$}
    \label{fig:1a}
  \end{subfigure}
  \begin{subfigure}[b]{0.23\textwidth}
    \includegraphics[width=\textwidth]{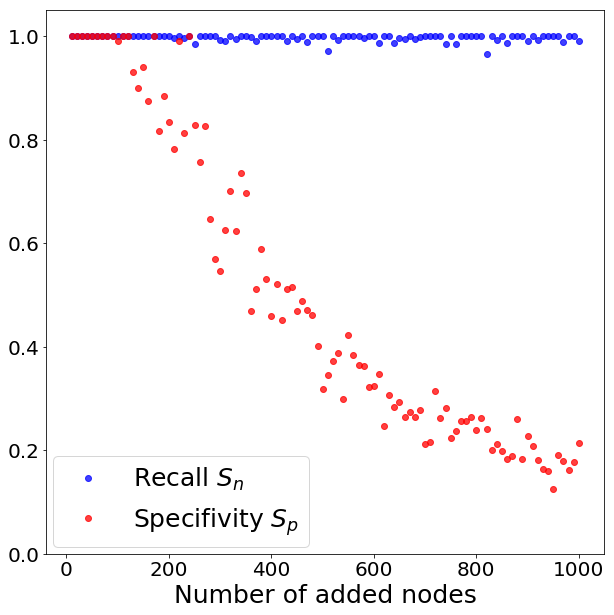}
    \caption{OSLOM $r=0$}
    \label{fig:2a}
  \end{subfigure}
  \begin{subfigure}[b]{0.23\textwidth}
    \includegraphics[width=\textwidth]{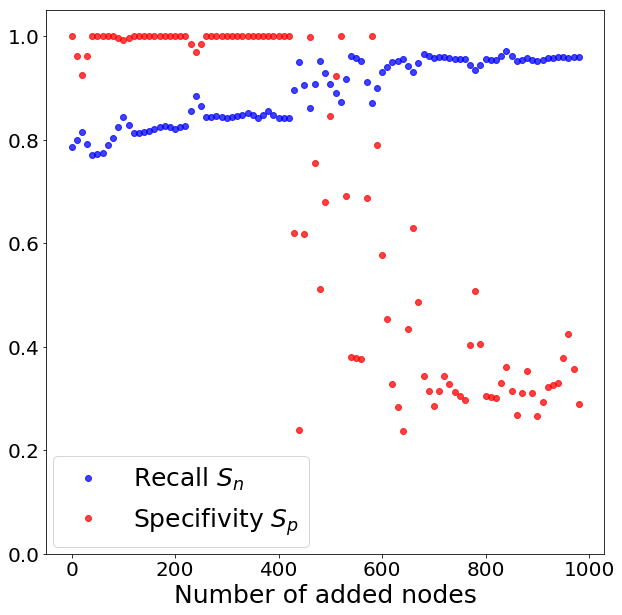}
    \caption{ItRich $r=1$}
    \label{fig:3a}
  \end{subfigure}
  \begin{subfigure}[b]{0.23\textwidth}
    \includegraphics[width=\textwidth]{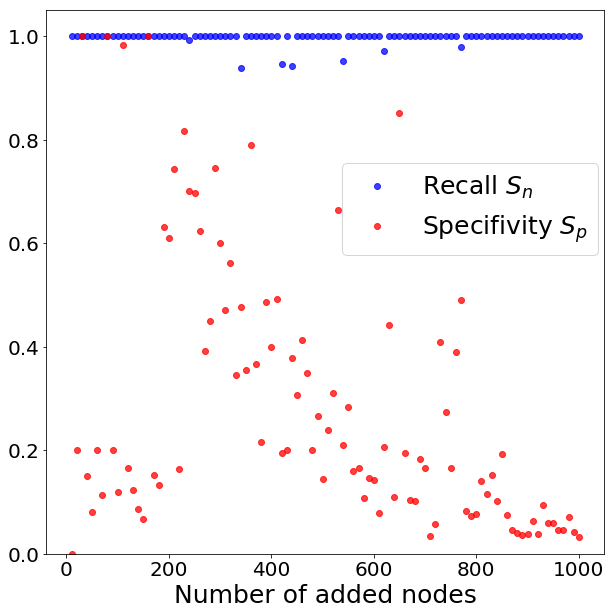}
    \caption{OSLOM $r=1$}
    \label{fig:4a}
  \end{subfigure}
  \caption{Values of ItRich and OSLOM performance measurements against the number of added nodes. For all cases $\mu=0.1$, $\beta=2,\gamma=3 $.  When $r=0$ (resp. $r=1$) the gap between the values of $\delta$ for the nodes of the LFR network and the added nodes is the highest (resp. null).}
    \label{ItRichvsOSLOM}
\end{figure}
We confirm the results of \cite{lan11} that OSLOM correctly separates the clusters and the noise as long as the noise is not too significant. ItRich is much more efficient to separate the LFR from the noise, even when the noise has a high density of links ($r=1$). 

When $r=0$, ItRich provides a Specificity equal to $1$ and that remains constant as we increase the number of added nodes. This means that all these nodes have been correctly classified in the sparse part, regardless of their number which ranges from $0$ to $1000$. Except for a few outliers, this is also the case when $r=1$ and the number of added nodes is less than half the number of the LFR nodes. On the contrary, in the OSLOM algorithm, the value of Specificity drops significantly and continuously after about $200$ nodes have been added, no matter whether $r=0$ or $r=1$. This suggests that the larger the number of added nodes, the more often some of them are assigned  to a community by OSLOM, resulting in a lower ratio of added nodes correctly assigned to the sparse part.

As for the Recall $Sn$, its values calculated by ItRich are high but less than $100\%$ from the beginning to the end of the calculation, even when there are no added nodes at all and whatever $r=0$ or $r=1$. This is due to the fact that there are some nodes within the LFR network that are initially classified as sparse, and that remains so when noise is added. 
In comparison,  OSLOM also keeps a constant value of $Sn$ with an average of $99$\%, and a standard deviation of 1\%. This shows that this algorithm  assigns a community to almost all nodes of the LFR network, even when $r=1$. 

Finally, two phenomena may be noted. The first is for OSLOM when $r=1$ and around 200 nodes have been added. There is a threshold effect that causes a jump in the values of the Specificity, from values lower than 0.2 to values around 0.8. This is explained by the fact that as long as the number of added nodes is small enough, OSLOM incorrectly groups them into a single community. The second is for ItRich when $r=1$ and around 500 nodes have been added. The Specificity seems to take random values between 0 and 1 and the Recall between 0.85 and 0.95. We have no clear explanation for this. 

\vspace{.3cm}
Concerning the experiments for $\mu=0.2$ and $\mu=0.5$ and 9 other values of $r$ between $0$ and $1$ (see Appendix 2), 
the best results of ItRich (for these three values of $\mu$) are obtained for $\mu=0.5$, and this can be accounted for by the fact that they correspond to the case in which the perturbations are the lowest. Indeed, by increasing the mixing parameter $\mu$, we decrease the average value of $\delta$ in the initial LFR network, which decreases the value of the gap $g_0$ and allows it to be filled without adding too many random links (i.e. the variance between the networks generated from $r=0$ to $r=1$ decreases with $\mu$). 
 We generally observe better results on average for ItRich than for OSLOM, especially regarding the Specificity measure, which has a minimum average value of $0.63 $ for ItRich when $\mu = 0.1$, $r=0.6$. It means that, in the worst case, our algorithm manages to identify on average at least $63\%$ of the ``noise" nodes. On the other hand, the Recall is better for OSLOM, but it often means misclassifying the added nodes in the dense part with the nodes of the initial LFR.

\subsection{Real world networks}\label{real-network}
We apply ItRich to three data sets that are widely studied in the field of network science: Lusseau's bottlenose dolphins \cite{Lusseau,lus04}, American political blogs \cite{Adamic2005} and American College football \cite{Girvan2002}. The first one has only 62 nodes and can thus be analyzed finely. The second one is a bigger network with 1490 nodes. Finally, the third one has all of its 115 nodes in a 8-shell except one \footnote{All the datasets have been downloaded from https://www.cc.gatech.edu/dimacs10/archive/clustering.shtml}. For all these networks, we choose the default value  $Q_{threshold}=Q_1/10$. 

\begin{figure}
  \begin{subfigure}[b]{0.31\textwidth}
    \includegraphics[width=\textwidth]{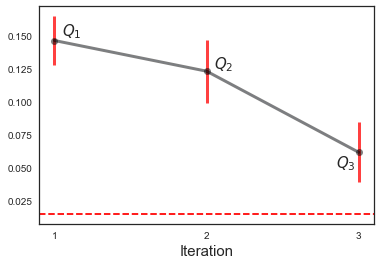}
    \caption{Dolphins}
    \label{fig:qualitiesa}
  \end{subfigure}
  \begin{subfigure}[b]{0.31\textwidth}
    \includegraphics[width=\textwidth]{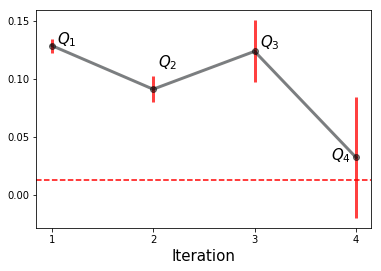}
    \caption{College football}
    \label{fig:qualitiesb}
  \end{subfigure}
    \begin{subfigure}[b]{0.31\textwidth}
    \includegraphics[width=\textwidth]{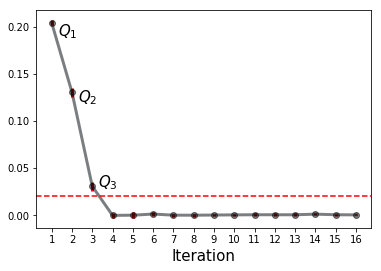}
    \caption{Political blogs}
    \label{fig:qualitiesc}
  \end{subfigure}
  \caption{The values of $Q$ for the three studied networks. The red dotted line represents the value of $Q_{threshold}=Q_1/10$. For each network, the null model is computed 100 times and the error bars represent the standard deviations of the quality measure. }
    \label{fig:qualities}
\end{figure}

\subsubsection{Lusseau's bottlenose dolphins}

The dolphin network is constructed from observations of a community of 62 bottlenose dolphins over a period of 7 years. Nodes in the network represent the dolphins, and ties between nodes represent associations between dolphin pairs occurring more often than expected by chance. 

\begin{figure}
   \begin{subfigure}[b]{0.49\textwidth}
    \includegraphics[width=\textwidth]{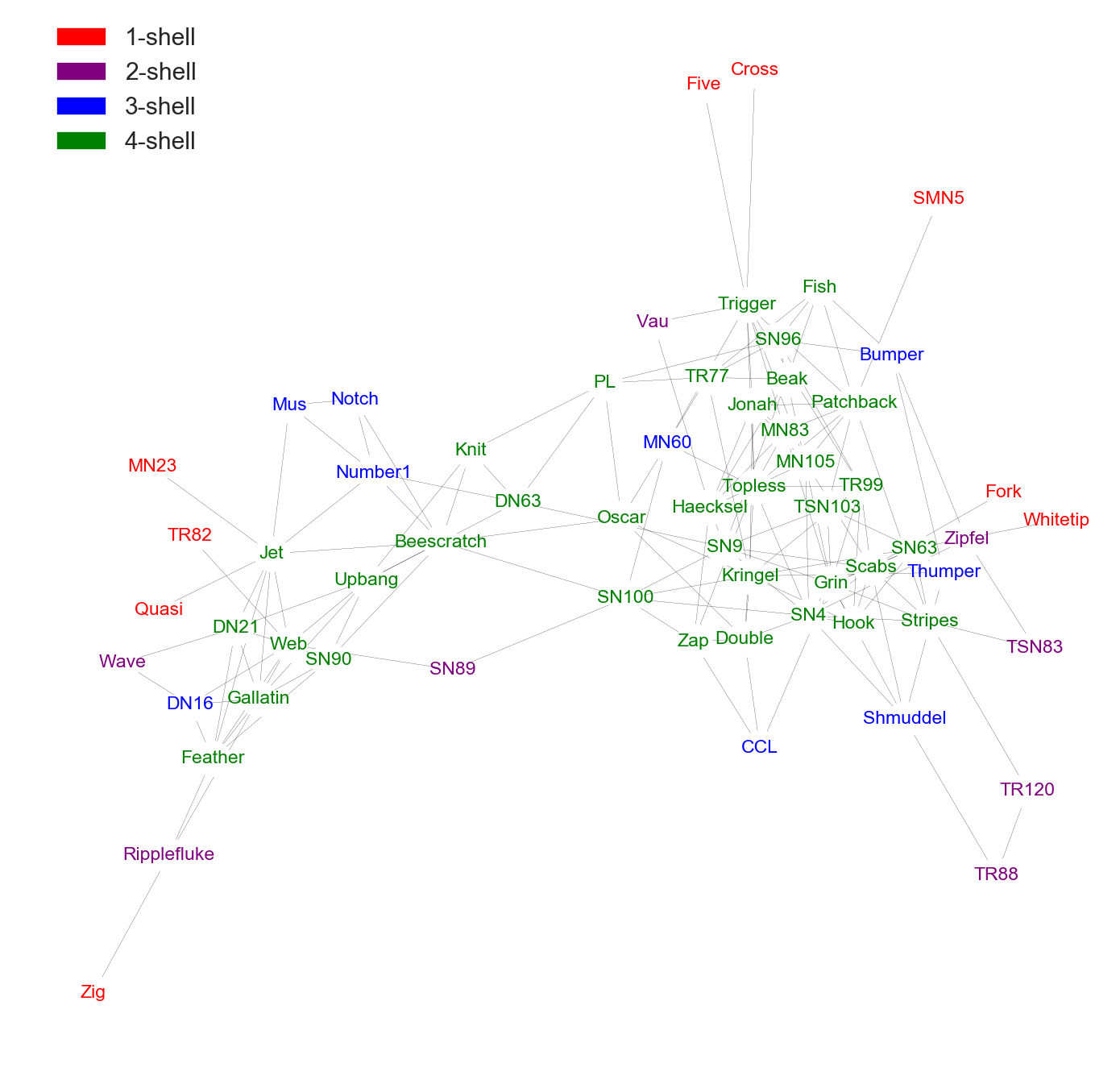}
    \caption{}
    \label{kardolph-a}
  \end{subfigure}
  \begin{subfigure}[b]{0.49\textwidth}
    \includegraphics[width=\textwidth]{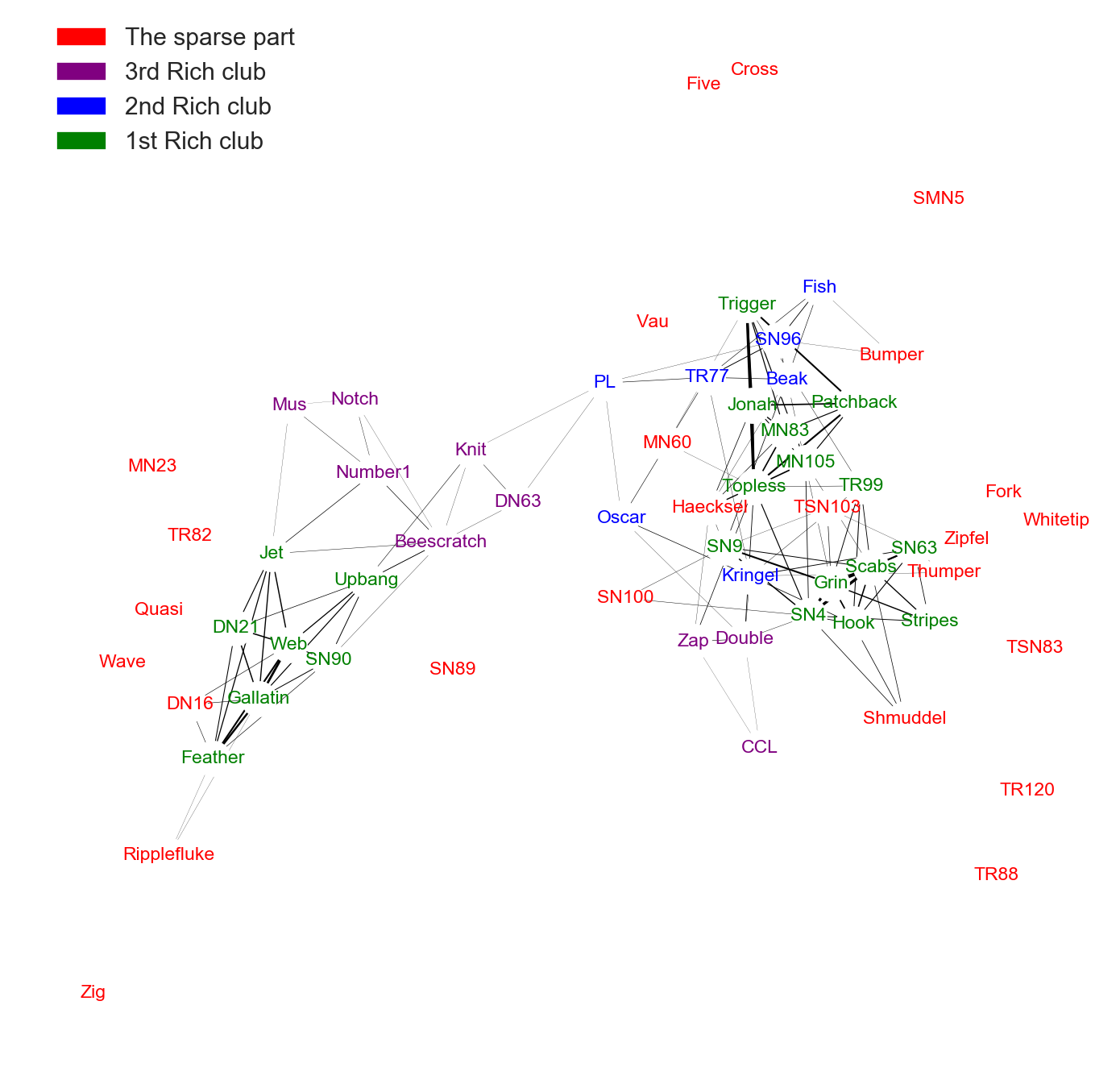}
    \caption{}
    \label{kardolph-b}
  \end{subfigure}
\caption{ Lusseau's bottlenose dolphins (a) The $k-$shell decomposition of the network. (b) The decomposition obtained from ItRich, the width of each edge is proportional to its weight $w$. The layouts are obtained with a force-based algorithm.}
\end{figure}

In \Cref{kardolph-b}, green, blue and purple nodes correspond respectively to the first, second and third weighted rich clubs identified by our algorithm. These results, compared to the community structure proposed in \cite{lus04} (\Cref{kardolph-a,kardolph-b}), illustrate the difference between ItRich and a community detection algorithm. Please note that ItRich may put the nodes of different communities in the same weighted rich club.

 We observe that the first weighted rich club is composed of highly connected nodes. The second and third weighted rich clubs are composed of several linked vertices mostly belonging to the neighbourhood of the first weighted rich club. These observations invite us to compare the distribution of the vertices within the weighted rich clubs and the distribution of the vertices within the $k-$shells \cite{k_core_decomp}. There are $4$ different shells in the dolphin network \Cref{kardolph-a}).  The sparse part contains the periphery which is equal to the 2-core. Symmetrically, the dense part is included in the union of the 3- shell and the 4- shell. 

Among the $62$ vertices of the network, $25$ are in the sparse part. These $25$ vertices are of two different types: those that have a null value of $\delta$ and those whose value of $\delta$ sharply decreases at each iteration of ItRich so that it remains lower than the minimal value required to be in a weighted rich club. These second-type vertices are Ripplefluke, MN60, SN100, TSN103, DN16, Shmuddel, Haecksel, Thumper, Bumper (\Cref{kardolph-a}). The first $6$ among them get a null value of $\delta$ as soon as the first weighted rich club is removed. Vertices TSN103, SN100 and Haecksel are in the 4-shell. Vertex TSN103 has $4$ neighbors, each of them of high degree but weakly linked together. In fact, TSN103 is at the intersection of two sets of vertices of two different communities, and so, has a very particular intermediary position. Vertex SN100 is the vertex with the highest betweenness in the network and the smallest non-zero clustering. It is also in a central position between distinct groups of vertices and has a relatively high degree. Lastly, Haecksel has a $\delta$ value that gradually decreases after removing the first and the second weighted rich clubs. It is largely linked to vertices of the first weighted rich club, which, when removed after the first iteration of ItRich, leads to a configuration in which Haecksel has a null clustering. So, within the 4-shell, the position of Haecksel is somewhat special.      

This first real world network study suggests that, when several k-cores exist, there may be a high overlap between the sparse (resp. dense) part and the k-cores for low (resp. high) values of $k$. The differences between both decompositions are worth looking at closely.  Moreover, the vertices of the sparse part of a network can be divided into two categories: those with a low $\delta$ value that are, from the beginning, at the periphery of the network and the others. The latter show medium values of $\delta$ but are linked to high-$\delta$ vertices. These vertices seem to occupy a rather specific position in the organisation of the core of the network.
Such an information is new compared to the one obtained from the analyses conducted so far, and it could lead to more in-depth interpretations by biologists.
%%%%%%%%%%%%%%%%%%%%%%%%%%%%%%%
\subsubsection{American political blogs}

This dataset is composed of a set of 1490 nodes representing American blogs discussing political issues. Each blog has been labeled by Lada Adamic \cite{Adamic2005} as either liberal (758) or conservative (732) and two blogs are linked if at least one refers to the other. The authors concluded that most links are within the two separate communities, with far fewer cross-links between them. Another interesting pattern was that conservative bloggers were more likely to link to other blogs (other conservative blogs but also liberal ones).

\medskip

For $Q_{threshold}=Q_1/10$, the dense part has three rich clubs which account $37\%$ of the nodes and $79\%$ of the links of the total graph. The distribution of nodes and links in the various rich clubs and the sparse part is given in \Cref{RC-Am-blog}. The three Rich clubs have comparable numbers of nodes and, within each rich club, the density of links between the liberal nodes on the one hand and between the conservative nodes on the other are almost equal. However, the 1st Rich club has a higher proportion of liberals, while the reverse is true for the other two rich clubs and for the sparse part.

\medskip
\begin{table}[]
    \centering
    \begin{tabular}{l|c|c|c}
    & liberal & conservative & total\\
    \hline
    Rich club 1   & 113/3052 ~(48.2)  & 84/1682 ~(48.2)& 197/5260 ~(27.2)\\
    Rich club 2     & 80/283 ~(9.0) &134/710 ~(8.0) & 214/1083 ~(4.8) \\
    Rich club 3   &50/42 ~(3.4)  &96/116 ~(2.5)  & 146/165 ~(1.6) \\
    Sparse part     &515/84~(0.06)  &418/50~(0.05) & 933/152~(0.03) \\
    \hline
    Total & 758/7302~(2.5) & 732/7841~(2.9) & 1490/16718/~(1.5)
    \end{tabular}
    \caption{Distribution of nodes and links within the American political blogs network: $A/B~(C)$ is for $A$ nodes, $B$ links and $C$ the density of links of the induced subgraph (ie. $C=2*\frac{B}{A*(A-1)}*100$). The links are not directed, that is, one link corresponds to an arc from a node $i$ to a node $j$, or from $j$ to $i$, or to a pair of arcs with opposite directions between $i$ and $j$.}
    \label{RC-Am-blog}
\end{table}

\begin{figure}[!h]
    \centering
   % ce symbole ajoute un espacement horisontal entre les premiÃ¨res deux images
    \begin{subfigure}[b]{0.45\textwidth}
        \centering \includegraphics[width=\textwidth]{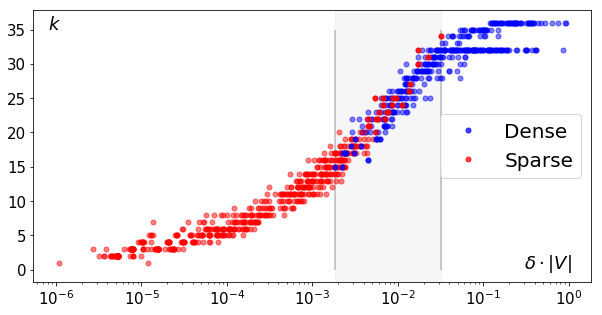}
        \caption{ }\label{fig:blog2-a}
    \end{subfigure}
       ~ 
 \begin{subfigure}[b]{0.45\textwidth} % "0.45" donne ici la largeur de l'image
        \centering \includegraphics[width=\textwidth]{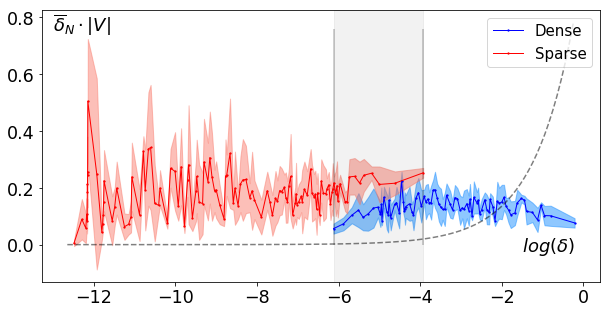}
        \caption{ }\label{fig:blog2-b}
    \end{subfigure}
 
    %la ligne blanche correspond au retour Ã  la ligne aprÃ¨s le deuxiÃ¨me image
   \begin{subfigure}[b]{0.45\textwidth}
        \centering \includegraphics[width=\textwidth]{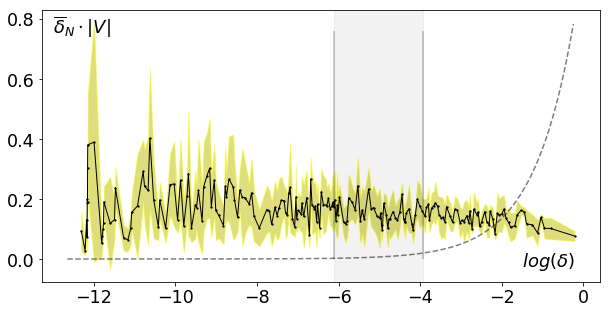}
        \caption{ }\label{fig:blog-c}
    \end{subfigure}
 %   ~ 
 %   \begin{subfigure}[b]{0.45\textwidth}
  %      \centering \includegraphics[width=\textwidth]{k_k_1.png}
 %       \caption{ }\label{fig:challenge1-d}
%\end{subfigure}
\caption{ American political blogs (a) Plot of $k$-core vs. $\delta$ with nodes of the dense (resp. sparse) part in blue (resp. red). (b,c) The average $\overline{\delta}_N$ of $\delta$ calculated on the neighbourhood of the nodes vs. their $\delta$ value on a logarithmic scale. For a given $\delta$, only 
the mean (bold curve) and standard deviation are given. In (b), the nodes of the dense and sparse parts are distinguished, which is not the case in (c). 
The overlapping area between dense and sparse parts is shown by vertical lines.}
\label{fig:quant}
\end{figure}

As in the previous data set, we observe a correlation between the $k$-core and the values of $\delta$, with an overlap of the values for the dense and the sparse parts (\Cref{fig:blog2-a}). However, the large number of $k$-cores (from the 15-core to the 34-core) that contains nodes of both the dense part and the sparse part does not make the $k$-core decomposition efficient to allow for the differentiation of the weighted rich clubs from the sparse part of the network. 

If we pay particular attention to the nodes whose $\delta$ is in the overlap of the $\delta$-values covered by both the dense and the sparse parts, \Cref{fig:blog2-b} confirms that the $\delta$-values of the neighbors of a node ($\overline{\delta}_N$) is of particular importance to differentiate the two types of nodes of the overlap: for a same $k$-shell, the nodes of the sparse part have neighbours with average $\delta$ values higher than those of nodes of the dense part. As many of these neighbours belong to one weighted rich club, the $\delta$ values of these vertices partially or totally collapse after removing some rich-clubs and this explains why, \textit{in fine}, they are categorised in the sparse part of the network. 

However, let us note that the knowledge of $\delta$ and $\overline{\delta}_N$, without using ItRich, is not sufficient to provide any information to characterize the weighted rich clubs or the sparse part, or even to find their overlap (\Cref{fig:blog-c}).    

\subsubsection{American College Football}
We will now examine a network representing the confrontations between different American college football teams during the 2000 season \cite{Girvan2002}. The nodes represent the participating teams and a link connects two teams which played against each other during the season. Each team is labeled by a conference that contains from 8 to 12 teams. For most conferences, internal matches are more frequent than external matches, giving the network a modular structure. We identify, however, two properties that make this data set particularly suitable for testing our algorithm. The first is that all the nodes are part of the $8-shell$, except one which is in the $7-shell$. This implies that the results cannot be found by a $k-$core decomposition. The second property is that there are 5 teams that are not part of any conference, which have been given the label ``independent''. In \cite{Girvan2002} it is also stated that the seven teams in the Sunbelt conference played almost as many games against teams in the Western Athletic conference as against teams in their own conference. They also played a large portion of their inter-conference games against teams from the Mid-American conference.

\begin{figure}
    \centering
    \includegraphics[width=\textwidth]{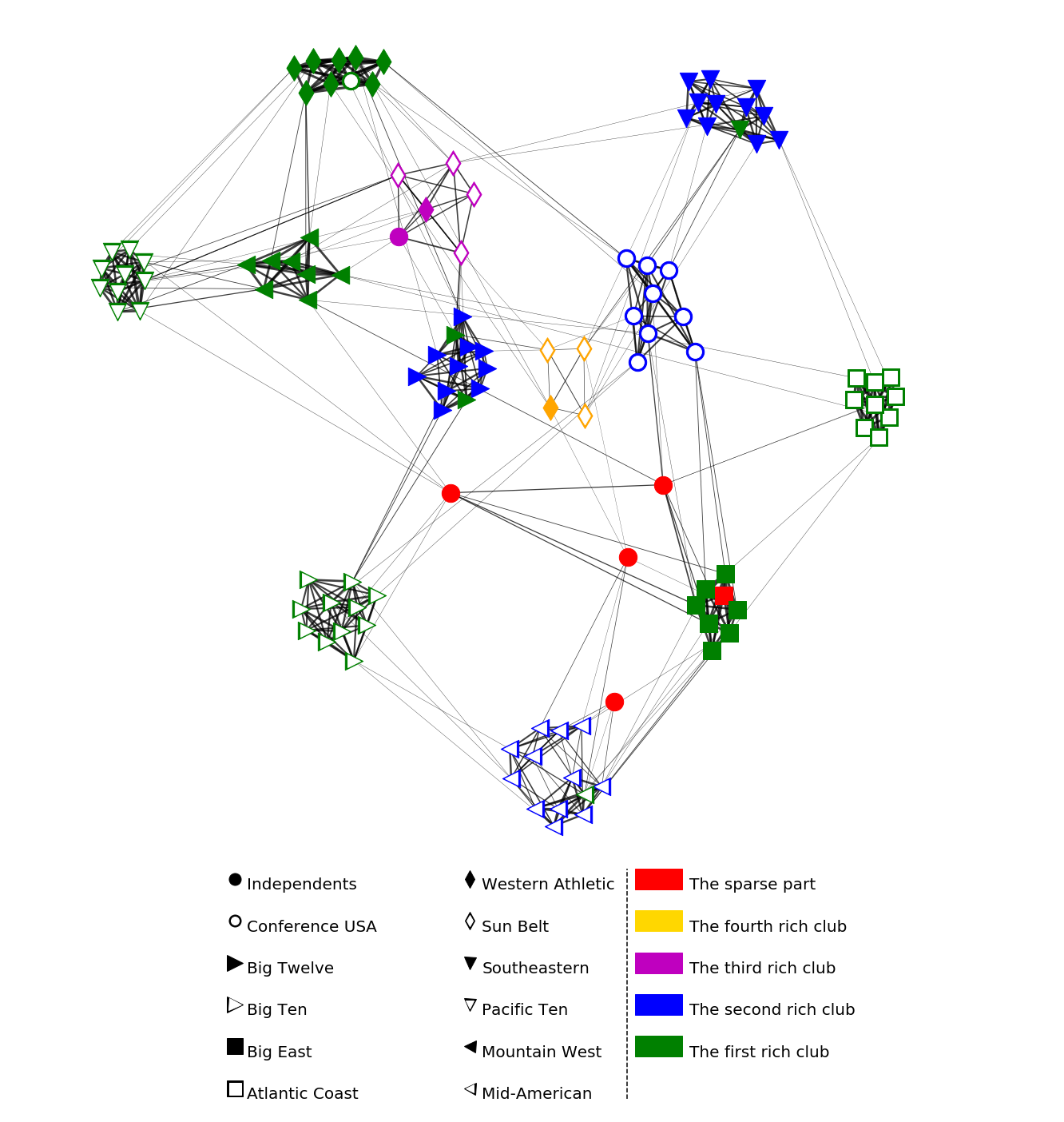}
    \caption{ American college football. In the decomposition obtained from ItRich, each node is represented by a different marker representing its conference, along with a different color according to its classification by Itrich. The width of each edge is proportional to its weight $w$. The layout is obtained with a force-based algorithm. }
    \label{fig:itrich_football}
\end{figure}

Figure ~\ref{fig:itrich_football} shows the rich club of each team and the conference it belongs to. 
ItRich reveals four rich clubs, the first two being composed of respectively 58 and 42 nodes, and the last two being smaller with respectively 6 and 4 nodes.  Five nodes remain in the sparse part.

It can be noted that out of the 4 rich clubs, the first two mainly contain teams that play the majority of their games against teams from their own conference, while the last two contain teams which tend to diversify their opponents' conferences (eg. the Sun Belt conference).
All but one of the teams of the first rich club are those correctly classified in \cite{Girvan2002} in the sense that the composition of the community it belongs to is exactly the composition of the conference it belongs to (Texas Christian is in the first rich club but misclassified in \cite{Girvan2002}). The last rich clubs contain teams that play a large number of inter-conference games, including teams from the Sun Belt conference, and some teams from the Western Athletic conference, which are, according to \cite{Girvan2002}, teams whose conference does not really form a community, in the sense that there are few intra-conference confrontations. We also notice that the third and the fourth rich clubs are equal to some communities found by Girvan and Newman's algorithm. This is explained by the fact that they both induce small cliques in the network (with 6 and 4 nodes respectively). 

To quantify the information carried by the links between teams and conferences, we use an empirical measure based on the Shannon's entropy.  Let $d(i)$ be the degree of the node $i \in V$, and $C = \{c_l\}_{l = 1,...,12}$ the set of the 11 conferences in which the teams play, plus the set of independent teams. For each node $i$ we call $p^{(i)}_l = \frac{|N(i) \cap c_l |}{d(i)}$ the ratio between the number of neighbours of the node $i$ playing in the conference $l$ and the total number of neighbours of $i$. 
We have 

\begin{equation}
    H(i) = -\sum_l p^{(i)}_l \cdot log(p^{(i)}_l)
\end{equation}

\begin{figure}
    \centering
    \includegraphics[width=.5\textwidth]{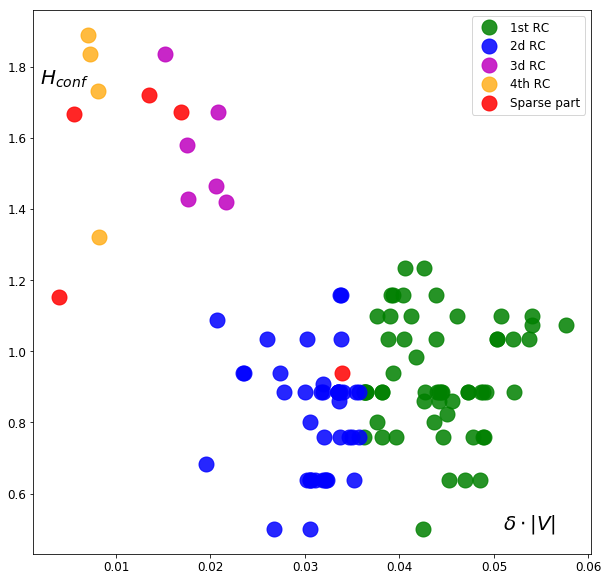}
    \caption{The quantities of information of each node versus its $\delta$ value. The nodes are distinguished by their colours according to the weighted rich club they belong to.  }
    \label{fig:entropy}
    
\end{figure}

This measure is zero if all the neighbours of $i$ play in the same conference, and has a maximum value of $\log12$ reached for a node when all its neighbours are equally distributed across the conferences. \Cref{fig:entropy} plots $H$ versus $\delta$. 
\newline The first two rich clubs are characterized by teams with low values of $H$ and high values of $\delta$. It reflects the fact that these teams mainly play intra-conference matches. On the contrary, teams of the last two rich clubs and those in the sparse part mainly play matches against teams of varied profiles.  
The sparse part is composed of 5 nodes, 4 of them (Navy, Central Florida, Notre Dame and Connecticut) are independent teams, and have not been classified among any community in \cite{Girvan2002}. The sparse part covers all the independent teams, except the Utah State team which is classified in the 3rd rich club, as it belongs to a clique. 

The only node in the sparse part that does not share these properties is the Miami Florida team (from the Big East conference), whose $\delta$ is high and $H$ is low. This node is in the situation described above, namely that, despite its high weight, it does not reach that of its neighbors, who are almost all in the first rich club. Remember that it is in the 8-core. 

\section{Conclusion}
This paper proposes a new viewpoint on the network structure analysis in order to provide both an alternative and an additional approach to standard methods, such as the k$-$core decomposition, or the community detection methods. We defined a new density measure for each vertex, called $\delta$, taking into account both the degree of the vertex and of its neighbors and the ratio of common neighbors between the vertex and its neighbors. We used this measure in the particular context of weighted rich clubs to develop an algorithm capable of providing several hierarchical layers of nodes, which altogether constitute what we call the ``dense part". The set of vertices that are not in any of the layers is called the ``sparse part". Experiments on both synthetic and real networks show that the dense part largely intersects with the $k$-core for a high enough $k$. The sparse part, on the other hand, contains peripheral nodes but also nodes of the core that have a special position in the core configuration and that are not screened by other methods. 
These vertices have properties different from those of their neighbors while not being on the periphery and often remaining within the core of the network. So, they constitute a kind of backbone of the network that meshes well with the partition into communities or into $k$-cores which helps to better understand the topological organization of the network. How to connect these nodes with vertices representatives of their community and with specific peripheral nodes is one of the themes of our future studies.
The time complexity of our algorithm ItRich can be reduced to $O(m\log m)$ using parallel computations. 
\section{Appendix}
\subsection{Appendix 1}
The purpose of this appendix is to give formal expressions of $\delta$ and clustering coefficient for the planted partition model (PPM) with a tree-like architecture used in \Cref{distribution of delta}. 
The PPM  we consider is a random network $G=(V,E)$ defined by:
\begin{itemize}
    \item a set of vertices $V=\{1,\ldots ,N\}$ partitioned into $r$ disjoint subsets ${C_{1},\ldots ,C_{r}}$ of $n$ vertices each;
    \item a matrix $P=p \mathbb{1}_{r \times r} + q B$ where $B$ is the adjacency matrix of a tree with $r$ nodes, and $(p,q)\in [0;1]$.  
\end{itemize}
In such a model, the $C_i$s are called communities or blocks. Matrix $B=(B_{ij})_{(i,j)\in \llbracket 1;r \rrbracket ^2}$ is the adjacency matrix of the tree linking the $r$ blocks and we say that two blocks $C_i$ and $C_j$ are adjacent if $B_{ij}=1$. Two nodes within the same block have a probability $p$ to share one edge, whereas this probability is $q$ between nodes of different but adjacent blocks and $0$ otherwise.
%$P=p \cdot  \mathbb{1}_{r \times r} + q \cdot B$, where $ \mathbb{1}_{r \times r} $ is an identity matrix and $B$ is a binary symmetric $r \times r$ matrix with zeros in the diagonal.
Since nodes of the same block share the same properties on average, in the following we refer to average values. For example we note $k_i$ the average degree of a node in the bloc $C_i$, and $K_i$ the number of blocks adjacent to the block $C_i$. We have:
  $$  k_i = n (p+qK_i) $$
The local clustering coefficient $\text{Clust}(i)$ of a node $i\in C_i$ is the ratio between the number of edges in its neighborhood and the number of pairs of vertices in its neighborhood.
It is easy to see that the number of pairs is  $\frac{k_i \cdot (k_i -1)}{2}\approx \frac{k_i^2}{2}$ for $np>>1$. 
In order to evaluate the number of edges $(j,k)$ in the neighborhood $N(i)$ of $i$, we consider $3$ disjoint cases: 
\begin{itemize}
    \item $j \in C_i$ and $k\in C_i$: their number is $\frac{(n-1)(n-2)}{2} p^3\approx \frac{n^2}{2} p^3$
    \item $j\in C_{i}$ and $k\notin C_{i}$: their number is $ n(n-1)pq^2K_i \approx n^2pq^2K_i$
    \item $j\in C_{j}\neq C_{i}$ and $k\in C_{j}$: their number is $ \frac{n(n-1)}{2}pq^2K_i \approx \frac{n^2}{2}pq^2K_i$
    %\item $j\in C_{j}\neq C_{i}$, $k\in C_{k}\neq C_{i}$: their number is $\frac{(r-1)(r-2)}{2}n^2q^3$
\end{itemize}
Since $B$ is the adjacency matrix of a tree, the three vertices cannot be located in three distinct blocks. 
It follows:
\begin{equation}\label{clust}
\text{Clust}(i)\approx p\cdot \frac{p^2 + 3q^2K_i }{(p+qK_i)^2}  
\end{equation}
To calculate $\delta(i)$, observe that 
$$
\delta(i)= \sum_{j \in N(i)}w(i,j)=\sum_{j \in N(i)\cap C_i}w(i,j)+\sum_{j \in N(i)\setminus C_i}w(i,j).
$$
Remember that $w(i,j) = \frac{1}{(N-1)^2(N-2)}\cdot \vert N(i) \cap N(j) \vert \cdot H(k_i,k_j) $, where $H(k_i,k_j)$ is the harmonic mean of the degrees $k_i$ and $k_j$. It is easy to verify that $H(k_i,k_j) = n(p+qK_i)(p+qK_j)/(p+q(\frac{K_i+K_j}{2}))$.

Let $j\in N(i)$:

\begin{itemize}
    \item if $j\in C_i$: there are $(n-2)p^2 \approx np^2$ common neighbors of $i$ and $j$ in $C_i$ and $nq^2K_i$ common neighbors of $i$ and $j$ not in $C_i$. Moreover, as $k_i=k_j$, we have $H(k_i,k_j) = n(p+qK_i)$. So, given $j_0 \in N(i) \cap C_i$, we have:
    $$\sum_{j \in N(i)\cap C_i}w(i,j)= \vert N(i)\cap C_i \vert \cdot w(i,j_0) \approx \frac{1}{r^3}\cdot p(p^2+q^2K_i)(p+qK_i)$$ by noticing that $N=nr$.
    \item if $j \in C_j \neq C_i$: there are $2(n-1)pq \approx 2npq$ common neighbors of $i$ and $j$ that are in $C_i$ or $C_j$ and there cannot be common other neighbors (ie. neither in $C_i$ nor in $C_j$) since $B$ is the adjacency matrix of a tree.  %$nq^2K_{i,j}$ common neighbors of $i$ and $j$ that are neither in $C_i$ nor in $C_j$ where $K_{i,j}$ is the number of blocks adjacent to both $C_i$ and $C_j$. 
    $$\sum_{j \in N(i)\setminus C_i}w(i,j)=\frac{2(n-1)pq}{(N-1)^2(N-2)}\sum_{j \in N(i)\setminus C_i}H(k_i,k_j) \approx \frac{2pq^2}{nr^3}\sum_{j}B_{ij}H(k_i,k_j)$$
    
   % Therefore, given $j_1 \in N(i) \setminus C_i$, we have:
    %$$\sum_{j \in N(i)\setminus C_i}w(i,j)=\vert N(i)\setminus C_i \vert \cdot w(i,j_1) \approx \frac{1}{r^3}\cdot \frac{2pq^2K_i(p+qK_i)(p+qK_j)}{p+q(\frac{K_i+K_j}{2})}$$
\end{itemize}
To summarize, 
\begin{equation}\label{delta}
\delta(i) \approx \frac{p(p+qK_i)}{r^3}\cdot \left(p^2+q^2K_i+2q^2 \sum_jB_{ij}\frac{p+qK_j}{p+q(\frac{K_i+K_j}{2})}\right).
\end{equation}
\subsection{Appendix 2}
The following plots show the results of ItRich and OSLOM for different values of the mixing parameter $\mu$ and  of the gap parameter $r$. See \Cref{LFR} for details on the experimental protocol. 

Each of the 11 points of the curves is an average over 100 iterations of the algorithm (ItRich or OSLOM) with a varying number of added nodes. 
The experiment is carried for $\beta=2$ and $\gamma=3$.
\begin{center}
\begin{figure}[h]
  \begin{subfigure}[b]{0.31\textwidth}
    \includegraphics[width=\textwidth]{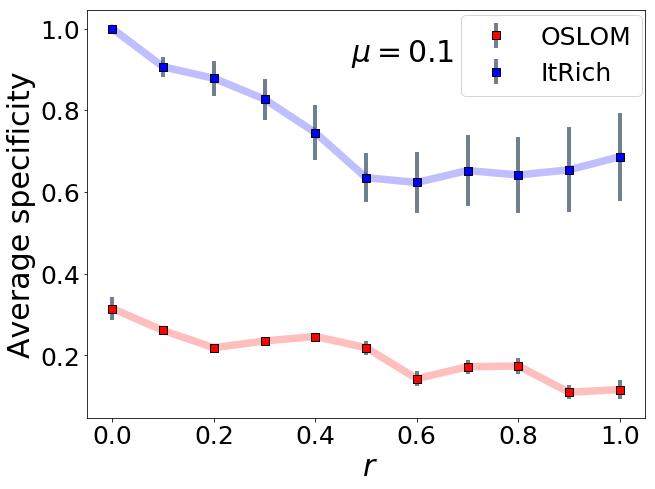}
    \caption{}
    \label{result-a}
  \end{subfigure}
  \begin{subfigure}[b]{0.31\textwidth}
    \includegraphics[width=\textwidth]{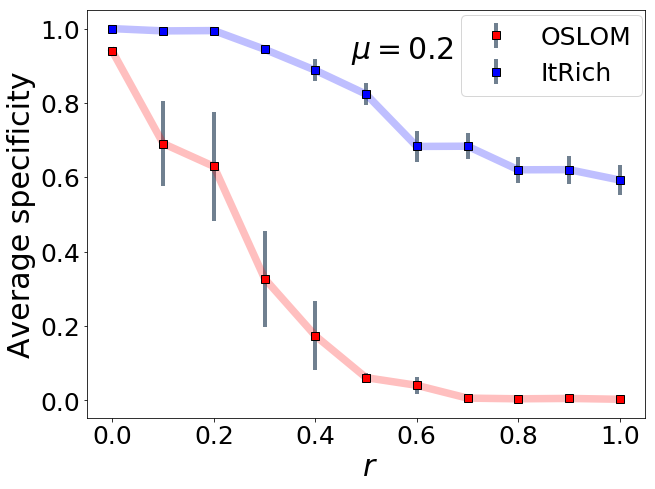}
    \caption{}
    \label{result-d}
  \end{subfigure}
  \begin{subfigure}[b]{0.31\textwidth}
    \includegraphics[width=\textwidth]{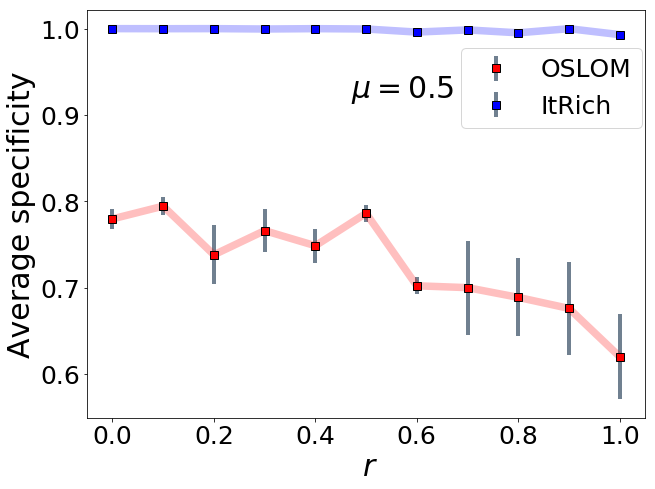}
    \caption{}
    \label{result-g}
  \end{subfigure}
  \begin{subfigure}[b]{0.31\textwidth}
    \includegraphics[width=\textwidth]{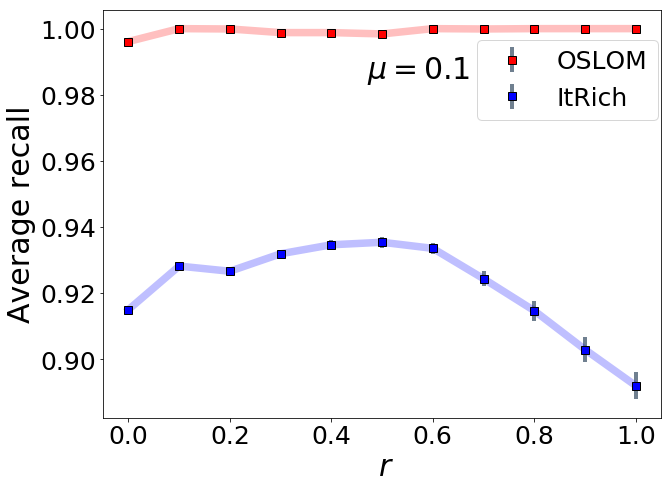}
    \caption{}
    \label{result-b}
  \end{subfigure}
  \begin{subfigure}[b]{0.31\textwidth}
    \includegraphics[width=\textwidth]{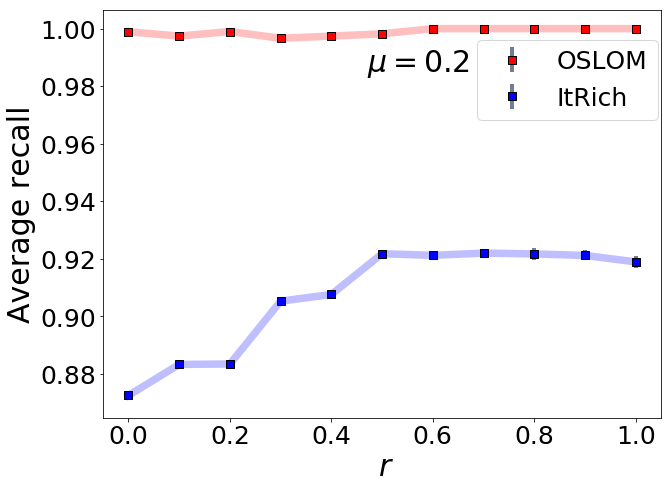}
    \caption{}
    \label{result-e}
  \end{subfigure}
  \begin{subfigure}[b]{0.31\textwidth}
    \includegraphics[width=\textwidth]{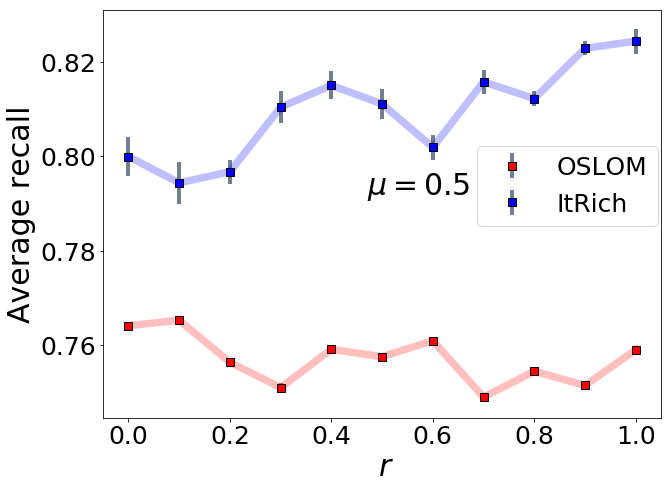}
    \caption{}
    \label{result-h}
  \end{subfigure}
   %
    %\begin{center}
%  \begin{subfigure}[b]{0.3\textwidth}
%    \includegraphics[width=\textwidth]{avg_F1_5.png}
 %   \caption{}
 %   \label{result-i}
 % \end{subfigure} 

\label{result}
\end{figure}
\end{center}
\bibliographystyle{imaiai}

\begin{thebibliography}{99}


\bibitem{Adamic2005}
\textsc{Adamic, L.~A.  {\small \&} Glance, N.}  (2005) The political
  blogosphere and the 2004 {U.S.} election: divided they blog. pp. 36--43.

\bibitem{alb99}
\textsc{Albert, R.~Jeong, H.  {\small \&} Barab\'{a}si, A.~L.}  (1999)
  Internet: Diameter of the World-Wide Web. \emph{Nature}, \textbf{401},
  130--131.

\bibitem{alb02}
\textsc{Albert, R.  {\small \&} Barab\'asi, A.~L.}  (2002) Statistical
  mechanics of complex networks. \emph{Reviews of Modern Physics}, \textbf{74},
  47--97.

\bibitem{als14}
\textsc{Alstott, J., Panzarasa, P., Rubinov, M., Bullmore, E.  {\small \&}
  V\'egrtes, P.~E.}  (2015) A unifying framework for measuring weighted rich
  clubs by integrating randomized controls. \emph{Scientific Reports},
  \textbf{4}, 7258.

\bibitem{bar99}
\textsc{Barab\'asi, A.~L.  {\small \&} Albert, R.}  (1999) Emergence of scaling
  in random networks. \emph{Science}, \textbf{286}, 509--512.

\bibitem{bia09}
\textsc{Bianconi, G., Pin, P.  {\small \&} Marsili, M.}  (2009) Assessing the
  relevance of node features for network structure. \emph{Proceedings of the
  National Academy of Sciences}, \textbf{106}, 11433--11438.

\bibitem{bol80}
\textsc{Bollob\'as, B.}  (1980) A probabilistic proof of an asymptotic formula
  for the number of labelled regular graphs. \emph{Eur. J. Comb.}, \textbf{1},
  311--316.

\bibitem{bor99}
\textsc{Borgatti, S.~P.  {\small \&} Everett, M.~G.}  (2000) Models of
  core/periphery structures. \emph{Social Networks}, \textbf{21}(4), 375 --
  395.

\bibitem{bou08}
\textsc{Boulet, R., Jouve, B., Rossi, F.  {\small \&} Villa, N.}  (2008) Batch
  kernel SOM and related Laplacian methods for social network analysis.
  \emph{Neurocomputing}, \textbf{71}(7), 1257--1273.

\bibitem{bur92}
\textsc{Burt, R.~S.}  (1992) \emph{Structural Holes: The Social Structure of
  Competition}. Cambridge: Harvard University Press.

\bibitem{che18}
\textsc{Chen, S., Wang, Z.~Z., Tang, L., Tang, Y., Gao, Y., Li, H.~J., Xiang,
  J.  {\small \&} Zhang, Y.}  (2018) Global vs. local modularity for network
  community detection. \emph{Plos One}, \textbf{13}, 1--21.

\bibitem{col06}
\textsc{Colizza, V., Flammini, A., Serrano, M.~A.  {\small \&} Vespignani, A.}
  (2006) Detecting rich-club ordering in complex networks. \emph{Nature
  Physics}, \textbf{2}, 110--115.

\bibitem{cre17}
\textsc{Crespelle, C.}  (2017) Structures of complex networks and of their
  dynamics. Habilitation \`a Diriger des Recherches, University of Lyon 1.

\bibitem{cse13}
\textsc{Csermely, P., London, A., Wu, L.~Y.  {\small \&} Uzzi, B.}  (2013)
  {Structure and dynamics of core/periphery networks}. \emph{Journal of Complex
  Networks}, \textbf{1}, 93--123.

\bibitem{dic45}
\textsc{Dice, L.~R.}  (1945) Measures of the amount of ecologic association
  between species. \emph{Ecology}, \textbf{26}, 297--302.

\bibitem{dor00}
\textsc{Dorogovtsev, S.~N.  {\small \&} Mendes, J. F.~F.}  (2003)
  \emph{Evolution of Networks: From Biological Nets to the Internet and WWW}.
  Oxford: Oxford University Press.

\bibitem{est11}
\textsc{Estrada, E.}  (2011) \emph{The Structure of Complex Networks: Theory
  and Applications}. Oxford University Press.

\bibitem{fal99}
\textsc{Faloutsos, M., Faloutsos, P.  {\small \&} Faloutsos, C.}  (1999) On
  power-law relationships of the Internet topology. \emph{SIGCOMM Comput.
  Commun. Rev.}, \textbf{29}(4), 251--262.

\bibitem{for16}
\textsc{Fortunato, S.  {\small \&} Hric, D.}  (2016) Community detection in
  networks: A user guide. \emph{Physics Reports}, \textbf{659}, 1--44.

\bibitem{fos18}
\textsc{Fosdick, B.~K., Larremore, D.~B., Nishimura, J.  {\small \&} Ugander,
  J.}  (2016) Configuring random graph models with fixed degree sequences.
  \emph{SIAM Review}, \textbf{60}, 315–355.

\bibitem{Girvan2002}
\textsc{Girvan, M.  {\small \&} Newman, M. E.~J.}  (2002) {Community structure
  in social and biological networks}. \emph{Proceedings of the National Academy
  of Sciences}, \textbf{99}(12), 7821--7826.


\bibitem{har99}
\textsc{Hartwell, L.~H., Hopfield, J.~J., Leibler, S.  {\small \&} Murray,
  A.~W.}  (1999) From molecular to modular cell biology. \emph{Nature},
  \textbf{402}, 47--52.

\bibitem{hol83}
\textsc{Holland, P.~W., Laskey, K.~B.  {\small \&} Leinhardt, S.}  (1983)
  Stochastic blockmodels: First steps. \emph{Social Networks}, \textbf{5}(2),
  109 -- 137.

\bibitem{hol05}
\textsc{Holme, P.}  (2005) Core-periphery organization of complex networks.
  \emph{Physical review. E, Statistical, nonlinear, and soft matter physics},
  \textbf{72}, 046111.

\bibitem{hri14}
\textsc{Hric, D., Darst, R.~K.  {\small \&} Fortunato, S.}  (2014) Community
  detection in networks: Structural communities versus ground truth.
  \emph{Physical review. E, Statistical, nonlinear, and soft matter physics},
  \textbf{90}, 1--19.

\bibitem{hub99}
\textsc{Huberman, B.~A.  {\small \&} Adamic, L.~A.}  (1999) Growth dynamics of
  the World-Wide Web. \emph{Nature}, \textbf{399}, 131.

\bibitem{jav18}
\textsc{Javed, A.~M., Younis, M.~S., Latif, S.~Qadir, J.  {\small \&} Baig, A.}
   (2018) Community detection in networks: A multidisciplinary review.
  \emph{Journal of Network and Computer Applications}, \textbf{108}, 87 -- 111.

\bibitem{jen16}
\textsc{Jensen, P., Morini, M., Karsai, M., Venturini, T., Vespignani, A.,
  Jacomy, M., Cointet, J.~P., Merckl\'e, P.  {\small \&} Fleury, E.}  (2015)
  Detecting global bridges in networks. \emph{Journal of Complex Networks},
  \textbf{4}(3), 319 -- 329.

\bibitem{kel12}
\textsc{Kelley, S., Goldberg, M., Magdon-Ismail, M., Mertsalov, K.  {\small \&}
  Wallace, A.}  (2012) Defining and discovering communities in social networks.
  in \emph{Handbook of Optimization in Complex Networks: Theory and
  Applications}, pp. 139--168. Boston: Springer US.

\bibitem{kra03}
\textsc{Krause, A.~E., Frank, K.~A., Mason, D.~M., Ulanowicz, R.~E.  {\small
  \&} Taylor, W.~W.}  (2003) Compartments revealed in food-web structure.
  \emph{Nature}, pp. 282--285.

\bibitem{kum10}
\textsc{Kumar, R., Novak, J.  {\small \&} Tomkins, A.}  (2006) Structure and
  evolution of online social networks. in \emph{Proceedings of the 12th ACM
  SIGKDD International Conference on Knowledge Discovery and Data Mining}, pp.
  611--617. New York: ACM.

\bibitem{kum99}
\textsc{Kumar, R., Raghavan, P., Rajagopalan, S.  {\small \&} Tomkins, A.}
  (1999) Trawling the Web for emerging cyber-communities. \emph{Comput. Netw.},
  \textbf{31}(11-16), 1481--1493.

\bibitem{lan09}
\textsc{Lancichinetti, A.  {\small \&} Fortunato, S.}  (2009) Benchmarks for
  testing community detection algorithms on directed and weighted graphs with
  overlapping communities. \emph{Phys. Rev. E}, \textbf{80}, 016118.

\bibitem{lan09a}
\textsc{Lancichinetti, A., Fortunato, S.  {\small \&} Kert{\'e}sz, J.}  (2009)
  Detecting the overlapping and hierarchical community structure in complex
  networks. \emph{New J. Phys.}, \textbf{11}(3), 033015.

\bibitem{lan08}
\textsc{Lancichinetti, A., Fortunato, S.  {\small \&} Radicchi, F.}  (2008)
  Benchmark graphs for testing community detection algorithms. \emph{Physical
  Review E}, \textbf{78}, 046110.

\bibitem{lan11}
\textsc{Lancichinetti, A., Radicchi, F., Ramasco, J.  {\small \&} Fortunato,
  S.}  (2011) Finding Statistically Significant Communities in Networks.
  \emph{Plos One}, \textbf{6}, e18961.

\bibitem{les10}
\textsc{Leskovec, J., Lang, K.~J.  {\small \&} Mahoney, M.}  (2010) Empirical
  comparison of algorithms for network community detection. in
  \emph{Proceedings of the 19th international conference on World wide web},
  pp. 631--640. New York: ACM.

\bibitem{LiuXu}
\textsc{Liu, B., Xu, S., Li, T., Xiao, J.  {\small \&} ke~X.}  (2018)
  Quantifying the Effects of Topology and Weight for Link Prediction in
  Weighted Complex Networks. \emph{Entropy}, \textbf{20}, 363.

\bibitem{lus04}
\textsc{Lusseau, D.  {\small \&} Newman, M. E.~J.}  (2004) Identifying the role
  that animals play in their social networks. \emph{Proceedings of the Royal
  Society B: Biological Sciences}, \textbf{271}, S477--S481.

\bibitem{Lusseau}
\textsc{Lusseau, D., Schneider, K., Boisseau, O.~J., Haase, P., Slooten, E.
  {\small \&} Dawson, S.~M.}  (2003) The bottlenose dolphin community of
  Doubtful Sound features a large proportion of long-lasting associations.
  \emph{Behavioral Ecology and Sociobiology}, \textbf{54}, 396--405.

\bibitem{meu10}
\textsc{Meunier, D., Lambiotte, R.  {\small \&} Bullmore, E.~T.}  (2010)
  Modular and Hierarchically Modular Organization of Brain Networks.
  \emph{Frontiers in Neuroscience}, \textbf{4}, 200.

\bibitem{mon12}
\textsc{Mones, E., Vicsek, L.  {\small \&} Vicsek, T.}  (2012) Hierarchy
  Measure for Complex Networks. \emph{Plos One}, \textbf{7}, e33799.

\bibitem{mor35}
\textsc{Moreno, J.~L.  {\small \&} Jennings, H.~H.}  (1938) Statistics of
  Social Configurations. \emph{Sociometry}, \textbf{1}, 342--374.

\bibitem{mor19}
\textsc{Moriya, S., Yamamoto, H., A.~H., Hirano-Iwata, A., Kubota, S.  {\small
  \&} Sato, S.}  (2019) Mean-field analysis of directed modular networks.
  \emph{Chaos: An Interdisciplinary Journal of Nonlinear Science},
  \textbf{29}(1), 013142.

\bibitem{new06}
\textsc{Newman, M., Barabasi, A.~L.  {\small \&} Watts, D.~J.}  (2006)
  \emph{The Structure and Dynamics of Networks}. Princeton University Press.

\bibitem{new03a}
\textsc{Newman, M. E.~J.}  (2003) The structure and function of complex
  networks. \emph{SIAM Review}, \textbf{45}, 167--256.

\bibitem{new10a}
\textsc{Newman, M. E.~J.}  (2010) \emph{Networks: An Introduction}. Oxford University
  Press.

\bibitem{new10}
\textsc{Newman, M. E.~J.}  (2016) Equivalence between modularity optimization and
  maximum likelihood methods for community detection. \emph{Phys. Rev. E},
  \textbf{94}, 052315.

\bibitem{new04}
\textsc{Newman, M. E.~J.  {\small \&} Girvan, M.}  (2004) Finding and
  evaluating community structure in networks. \emph{Phys. Rev. E}, \textbf{69},
  026113.

\bibitem{new03}
\textsc{Newman, M. E.~J.  {\small \&} Park, J.}  (2003) Why social networks are
  different from other types of networks. \emph{Phys. rev. E}, \textbf{68}(3),
  8.

\bibitem{ops08}
\textsc{Opsahl, T., Colizza, V., Panzarasa, P.  {\small \&} Ramasco, J.~J.}
  (2008) Prominence and Control: The Weighted Rich-Club Effect. \emph{Phys.
  Rev. Lett.}, \textbf{101}, 168702.

\bibitem{pee17}
\textsc{Peel, L., Larremore, D.~B.  {\small \&} Clauset, A.}  (2017) The ground
  truth about metadata and community detection in networks. \emph{Science
  Advances}, \textbf{3}(5), e1602548.

\bibitem{qua17}
\textsc{Qian, Y., Li, Y., Zhang, M., Ma, G.  {\small \&} Lu, F.}  (2017)
  Quantifying edge significance on maintaining global connectivity.
  \emph{Scientific Reports}, \textbf{7}, 45380.

\bibitem{rad04}
\textsc{Radicchi, F., Castellano, C., Cecconi, F., Loreto, V.  {\small \&}
  Parisi, D.}  (2004) Defining and identifying communities in networks.
  \emph{PNAS}, \textbf{101}, 2658--2663.

\bibitem{rav03}
\textsc{Ravasz, E.  {\small \&} Barab\'asi, A.~L.}  (2003) Hierarchical
  organization in complex networks. \emph{Phys. Rev. E}, \textbf{67}, 026112.

\bibitem{rav02}
\textsc{Ravasz, E., Somera, A.~L., Mongru, D.~A., Oltvai, Z.~N.  {\small \&}
  Barab{\'a}si, A.~L.}  (2002) Hierarchical organization of modularity in
  metabolic networks. \emph{Science}, \textbf{297}, 1551.

\bibitem{red98}
\textsc{Redner, S.}  (1998) How popular is your paper? An empirical study of
  the citation distribution. \emph{European Physical Journal B}, \textbf{4},
  131--134.

\bibitem{rom14}
\textsc{Rombach, M.~P., Porter, M.~A., Fowler, J.~H.  {\small \&} Mucha, P.~J.}
   (2014) {Core-Periphery Structure in Networks}. \emph{SIAM Journal on Applied
  Mathematics}, \textbf{74}, 167--190.

\bibitem{sch17}
\textsc{Schaub, M.~T., Delvenne, J.~C., Rosvall, M.  {\small \&} Lambiotte, R.}
   (2017) The many facets of community detection in complex networks.
  \emph{Applied Network Science}, \textbf{2}.

\bibitem{k_core_decomp}
\textsc{Seidman, S.~B.}  (1983) Network structure and minimum degree.
  \emph{Social Networks}, \textbf{5}, 269 -- 287.

\bibitem{sei13}
\textsc{Seifi, M., Junier, I., Rouquier, J.~B., Iskrov, S.  {\small \&}
  Guillaume, J.~L.}  (2013) Stable community cores in complex networks. in
  \emph{Complex Networks}, pp. 87--98. Berlin: Springer.

\bibitem{serrano06}
\textsc{Serrano, M., N\'{a}, M.~B.  {\small \&} Satorras, R.~P.}  (2006)
  Correlations in weighted networks. \emph{Physical Review E}, \textbf{74},
  055101.

\bibitem{ser08}
\textsc{Serrano, M.~A.}  (2008) Rich-club vs rich-multipolarization phenomena
  in weighted networks. \emph{Phys. Rev. E}, \textbf{78}, 026101.

\bibitem{sim62}
\textsc{Simon, H.}  (1962) The Architecture of Complexity. \emph{Proc. of the
  American Philosophical Society}, \textbf{106}, 467--482.

\bibitem{spi03}
\textsc{Spirin, V.  {\small \&} Mirny, L.~A.}  (2003) Protein complexes and
  functional modules in molecular networks. \emph{PNAS}, \textbf{100},
  12123--12128.

\bibitem{swi90}
\textsc{Swindale, N.~V.}  (1990) Is the cerebral cortex modular?. \emph{Trends
  in Neurosciences}, \textbf{13}, 487 -- 492.

\bibitem{ver14}
\textsc{Verma, T., Ara{\'u}jo, N. A.~M.  {\small \&} Herrmann, H.~J.}  (2014)
  Revealing the structure of the world airline network. \emph{Scientific
  Reports}, \textbf{4}, 5638.

\bibitem{Vijay}
\textsc{Vijaymeena, M.~K.  {\small \&} Kavitha, K.}  (2016) A Survey on
  Similarity Measures in Text Mining. \emph{Machine Learning and Applications:
  An International Journal}, \textbf{3}, 19--28.

\bibitem{wan11}
\textsc{Wang, Y., Di, Z.  {\small \&} Fan, Y.}  (2011) Identifying and
  Characterizing Nodes Important to Community Structure Using the Spectrum of
  the Graph. \emph{Plos One}, \textbf{6}(11), e27418.

\bibitem{was94}
\textsc{Wasserman, S.  {\small \&} Faust, K.}  (1995) \emph{Social network
  analysis: methods and applications}. Cambridge: University Press.

\bibitem{wat99}
\textsc{Watts, D.}  (1999) \emph{Small worlds: The dynamics of networks between
  order and randomness}. Princeton: Princeton University Press.

\bibitem{wat98}
\textsc{Watts, D.~J.  {\small \&} Strogatz, S.~H.}  (1998) Collective dynamics
  of ``small-world" networks. \emph{Nature}, \textbf{393}, 440--442.

\bibitem{xia18}
\textsc{Xiang, B.~B., Bao, Z.~K., Ma, C., Zhang, X., Chen, H.~S.  {\small \&}
  Zhang, H.~F.}  (2018) A unified method of detecting core-periphery structure
  and community structure in networks. \emph{Chaos: An Interdisciplinary
  Journal of Nonlinear Science}, \textbf{28}(1), 013122.

\bibitem{xia12}
\textsc{Xiang, J.  {\small \&} Hu, K.}  (2012) Limitation of multi-resolution
  methods in community detection. \emph{Physica A: Statistical Mechanics and
  its Applications}, \textbf{391}, 4995--5003.

\bibitem{yan12}
\textsc{Yang, J.  {\small \&} Leskovec, J.}  (2012) Community-Affiliation Graph
  Model for Overlapping Network Community Detection. in \emph{ICDM}, ed. by
  M.~J. Zaki, A.~Siebes, J.~X. Yu, B.~Goethals, G.~I. Webb,  {\small \&} X.~Wu,
  pp. 1170--1175. IEEE Computer Society.

\bibitem{you92}
\textsc{Young, M.~P.}  (1992) Objective analysis of the topological
  organization of the primate cortical visual system. \emph{Nature},
  \textbf{358}, 152--155.

\bibitem{Zachary}
\textsc{Zachary, W.~W.}  (1977) An information flow model for conflict and
  fission in small groups. \emph{Journal of Anthropological Research},
  \textbf{33}, 452--473.

\bibitem{zho04}
\textsc{Zhou, S.  {\small \&} Mondrag\`on, R.~J.}  (2004) The rich-club
  phenomenon in the Internet topology. \emph{IEEE Communications Letters},
  \textbf{8}, 180--182.

\bibitem{zla09}
\textsc{Zlatic, V., Bianconi, G., Diaz-Guilera, A., Garlaschelli, D., Rao, F.
  {\small \&} Caldarelli, G.}  (2009) On the rich-club effect in dense and
  weighted networks. \emph{The European Physical Journal B}, \textbf{67},
  271--275.


\end{thebibliography}

\end{document}